\documentclass[12pt,superscriptaddress]{revtex4}
\usepackage{latexsym, color, graphicx, comment}
\usepackage{amssymb}
\usepackage{amsmath}

\newcommand{\vect}[1]{\mbox{\boldmath $#1$}}
\newcommand{\tens}[1]{\vect{#1}}

\newcommand{\PS}{Pfirsch-Schl\"{u}ter~}

\newcommand{\vma}{\vect{v}_{\mathrm{m}a}}
\newcommand{\vm}{\vect{v}_{\mathrm{m}}}

\newcommand{\vE}{\vect{v}_{E}}

\newcommand{\vEo}{\vect{v}_{E0}}
\newcommand{\vpar}{v_{||}}

\newcommand{\vi}{v_{\mathrm{i}}}
\newcommand{\me}{m_{\mathrm{e}}}
\newcommand{\mi}{m_{\mathrm{i}}}
\newcommand{\Ti}{T_{\mathrm{i}}}

\newcommand{\Te}{T_{\mathrm{e}}}

\newcommand{\nuii}{\nu_{\mathrm{ii}}}

\newcommand{\sgn}{\mathrm{sgn}}
\newcommand{\Erres}{E_r^{\mathrm{res}}}
\newcommand{\Bpol}{B_{\mathrm{pol}}}
\newcommand{\Sap}{S_{a\mathrm{p}}}
\newcommand{\Sah}{S_{a\mathrm{h}}}
\newcommand{\changed}[1]{#1}

\begin{document}

\title{Comparison of particle trajectories and collision operators for collisional transport in nonaxisymmetric plasmas}



\author{M Landreman}
\email[]{mattland@umd.edu}
\affiliation{Institute for Research in Electronics and Applied Physics, University of Maryland, College Park, MD, 20742, USA}
\author{H M Smith}
\affiliation{Max-Planck-Institut f\"{u}r Plasmaphysik, 17491 Greifswald, Germany}
\author{A Moll\'{e}n}
\affiliation{Department of Applied Physics, 
  Chalmers University of Technology, 
  G\"oteborg,
  Sweden}
\author{P Helander}
\affiliation{Max-Planck-Institut f\"{u}r Plasmaphysik, 17491 Greifswald, Germany}


\date{\today}

\begin{abstract}

In this work, we examine the validity of several common simplifying assumptions used in numerical neoclassical calculations
for nonaxisymmetric plasmas,
both by using a new continuum drift-kinetic code
and by considering analytic properties of the kinetic equation.
First, neoclassical phenomena are computed for the LHD and W7-X stellarators using several versions of the drift-kinetic
equation, 
including the commonly used incompressible-$\vect{E}\times\vect{B}$-drift approximation
and two other variants,
corresponding to different effective particle trajectories.
It is found that
for electric fields below roughly one third of the resonant value, the different formulations give
nearly identical results, demonstrating the incompressible $\vect{E}\times\vect{B}$-drift approximation
is quite accurate in this regime. 
However, near the electric field resonance,
the models yield substantially different results.
We also compare results for various collision operators, including the full linearized Fokker-Planck operator.
At low collisionality, the radial transport driven by radial
gradients is nearly identical for the different operators,
while in other cases it is found to be important that collisions conserve momentum.

\end{abstract}


\maketitle 

\section{Introduction}

One important difference between axisymmetric and nonaxisymmetric plasmas is
that neoclassical effects in the latter are more sensitive to small values of the radial
electric field $E_r$. In axisymmetric plasmas, in order for the radial electric to modify the collisional ion heat flux
and other neoclassical phenomena,
the poloidal ion Mach number $(B / \Bpol) |\vect{v}_E|/\vi$ must approach $\sim 1$,
since an $E_r$ of corresponding magnitude is required to modify the trapped region of phase space\cite{GrishaNeo}.
Here, $B$ is the magnetic field magnitude, $\Bpol$ is the poloidal magnetic field,
$\vect{v}_E$ is the $\vect{E}\times\vect{B}$ drift, and $\vi=\sqrt{2 \Ti / \mi}$ is the ion thermal speed.
However, in nonaxisymmetric plasmas, 
a much smaller value of $E_r$ can modify the collisional fluxes \cite{Galeev, HoKulsrud, BeidlerBigBenchmarking}.
The reason is that helically trapped particles experience a secular radial magnetic drift, and
whichever process first interrupts this radial motion will thereby determine the step size for radial diffusion.
When $E_r=0$, the radial magnetic drift is interrupted by collisions, which cause
the particle to gain parallel momentum and de-trap.
But if $E_r$ is sufficient for the poloidal $\vect{E}\times\vect{B}$
precession frequency to exceed the effective collisional detrapping rate, $\vect{E}\times\vect{B}$ precession
begins to carry helically trapped particles onto untrapped trajectories, 
and also confines the trapped orbits by convecting them (usually poloidally) around the torus,
thereby limiting the radial step size and transport.
This transition from collisional ($1/\nu$-regime) to $E_r$-limited ($\sqrt{\nu}$-regime) transport
typically occurs at values of $E_r$ for which the poloidal Mach number is still $\ll 1$, due to the low collisionality
in typical experiments. (Here, $\nu$ denotes a collision frequency.)
For this reason, stellarator transport at low collisionality is sensitive to small values of $E_r$.
A variety of codes have been developed to compute 
these neoclassical effects in stellarators \cite{DKES1, DKES2, GSRAKE, Tribaldos2001, PENTA, Kernbichler1, FORTEC_PFR, FORTEC_CPC, JOSE}.

However, including the physics of $\vect{E}\times\vect{B}$ precession in a $\delta\! f$ drift-kinetic equation (or code to
solve such an equation)
is complicated by several issues.
First, if a rigorous expansion in $\rho_* \ll 1$ is employed, 
$\vect{E}\times\vect{B}$ precession is formally excluded
when the usual drift ordering $\vect{v}_E \sim \rho_* \vi$ is used,
but the high-flow ordering $\vect{v}_E \sim \vi$ is
not a useful ordering either, since it leads to contradictions in a general nonaxisymmetric field \cite{HelanderRotation, SugamaHighFlow}.
Here, $\rho_* = \rho/L$ where $\rho$ is the ion gyroradius and $L$ is a typical
macroscopic scale length.
Second, if the $\vect{v}_{E}$ poloidal precession term
is included in a radially local, time-independent kinetic equation
for $\delta\! f$ (the departure of the distribution function from a Maxwellian),
unphysical constraints are placed on the distribution function,
as we will prove in section \ref{sec:sources}
by considering appropriate moments of the kinetic equation.
These constraints only appear when $E_r \ne 0$, meaning a small but nonzero
$E_r$ is a singular perturbation of the $E_r=0$ case.
These unphysical behaviors have been eliminated in previous codes\cite{BeidlerBigBenchmarking}
by making the ad-hoc replacement $1/B^2 \to 1/\left< B^2 \right>$ (where $\left< \ldots \right>$ denotes
a flux surface average) in the $\vect{E}\times\vect{B}$ drift.
At the same time, variation in the particles' energy and pitch angle associated with $E_r$ is neglected.
These replacements and omissions are chosen so as to restore the variational form of the kinetic equation  \cite{DKES1, DKES2}.
These changes to the kinetic equation may be called the ``incompressible-$\vect{E}\times\vect{B}$-drift''
approximation \cite{Beidler}.
Some investigations have indicated that the incompressible-$\vect{E}\times\vect{B}$-drift
approximation may be reasonably accurate for small $E_r$ but a poor approximation
for larger $E_r$ \cite{Beidler, meMonoenergetic}.
This issue of which collisionless terms to include in the kinetic equation
is effectively a choice between particle trajectories, since the collisionless guiding center trajectories
are equivalent to the characteristic curves of the drift-kinetic equation.

Another limitation of many past stellarator neoclassical calculations
is that they are often performed with simplified models for collisions.
The linearized Fokker-Planck collision operator -- the most accurate linear operator available -- 
has been implemented in a variety of 
tokamak neoclassical codes \cite{Sauter0, Sauter, WongChan, NEOFP, speedGrids}.
However, due to the numerical challenge of the extra dimension in stellarators (i.e., the lack of toroidal symmetry), 
many stellarator neoclassical codes retain only pitch-angle scattering collisions,
so coupling in the energy dimension is eliminated.
The pitch-angle scattering operator
lacks the momentum conservation property of the Fokker-Planck operator,
which is known to be important in many situations \cite{Briesemeister}.
Several techniques have been devised and implemented \cite{Taguchi2, SugamaNishimura, Maassberg, PENTA}
to effectively restore momentum conservation by post-processing
the transport coefficients obtained with a pure pitch-angle scattering operator,
but these methods will not exactly reproduce calculations with the full linearized Fokker-Planck operator.
The NEO-2 code has implemented the full linearized Fokker-Planck operator for
stellarator geometry \cite{Kernbichler1}, but using a field-line-tracing method
which makes it difficult to add the important effect of poloidal $\vect{E}\times\vect{B}$ precession.

Here, we describe a new stellarator neoclassical code
SFINCS (the Stellarator Fokker-Planck Iterative Neoclassical Conservative Solver)
that can be used to explore the aforementioned issues,
comparing various models for effective particle trajectories and collisions.
Although we use the terminology of ``effective trajectories,'' the code
uses continuum rather than Monte Carlo algorithms.
The code solves the 4D drift-kinetic equation for the distribution function,
retaining coupling in 2 spatial independent variables (toroidal and poloidal angle)
and 2 velocity independent variables (speed and pitch angle),
but neglecting radial coupling.
(For comparison, DKES \cite{DKES1, DKES2} is 3D since energy coupling is neglected, 
while FORTEC-3D \cite{FORTEC_PFR, FORTEC_CPC} is 5D since
radial coupling is retained.)
General nonaxisymmetric nested flux surface geometry is allowed,
one or more species may be included, and several models for collisions are available,
including the full inter-species linearized Fokker-Planck operator.
The incompressible-$\vect{E}\times\vect{B}$-drift trajectories are implemented,
as are several other options for trajectories that include the true $\vect{E}\times\vect{B}$ drift.
As we shall demonstrate, retaining the true form of the $\vect{E}\times\vect{B}$
drift comes at a cost, requiring sources/sinks in the kinetic equation
in order for the solutions to be well behaved.
While all of the various options for the particle trajectories have disadvantages,
SFINCS allows the options to be compared.
As we will show in several calculations for the LHD and W7-X stellarators,
in many experimentally relevant cases,
the transport matrix elements are nearly identical for the various choices of particle trajectories.
However, differences between the trajectory models emerge when the radial electric field
grows comparable to the ``resonant'' value.

In the following section, we motivate the form of the kinetic equation solved by SFINCS,
and detail the three models for particle trajectories that will be compared.
For several of the particle trajectory models,
additional sources/sinks and constraints must be included in the system of equations
for the equations to be well posed and for the solutions to be well behaved.
These issues are explored in section \ref{sec:sources}.
In section \ref{sec:momentum}, we discuss some observations regarding momentum conservation,
and demonstrate that the electric field terms in the kinetic description
correspond to a component of gyroviscosity in a fluid description
only for the most accurate trajectory model.
Details of the numerical implementation are given in section \ref{sec:numerics}.
Some of the numerical results presented are given in terms of a transport
matrix, which is defined in section \ref{sec:transportMatrix}.
The numerical results are presented in sections \ref{sec:ErComparison} and \ref{sec:collisionComparison},
in which we discuss the transport matrix elements for the geometries of the LHD and W7-X
stellarators, comparing a variety of assumptions about the particle trajectories and collision operator.
In section \ref{sec:conclusions} we discuss the results and conclude.

\section{Kinetic equations}
\label{sec:equations}

We begin with the drift-kinetic equation (19) of Ref. \cite{Hazeltine}.
The standard drift ordering is applied at first: $\rho_{*a} \ll 1$ where $\rho_{*a} = \rho_a/L$,
$\vE/v_a \sim \rho_{*a}$, $\partial/\partial t \sim \rho_{*a}^2 v_a/L$, and $\nu_a \sim v_a/L$.
Here, $v_a = \sqrt{2 T_a / m_a}$ is the thermal speed of species $a$, $T_a$ is the temperature,
$m_a$ is the mass, $\rho_a = v_a m_a c/(Z_a e B)$ is the gyroradius, $Z_a$ is the species charge in units
of the proton charge $e$, $c$ is the speed of light, $L$ is a typical scale length,
and $\nu_a$ is a collision frequency.
No expansion in mass ratios or charges is made.
We expand the distribution function as $f_a = f_{a0} + f_{a1} + \ldots$.
The leading order distribution function $f_{a0}$ is taken to be a Maxwellian
that is constant on flux surfaces when expressed in terms of total energy
$W_a = v^2/2+Z_a e \Phi/m_a$:
\begin{equation}
f_{a0} = \eta_a(\psi) \left[ \frac{m_a}{2\pi T_a(\psi)}\right]^{3/2} \exp\left( -\frac{m_a W_a}{T_a(\psi)}\right).
\label{eq:Maxwellian}
\end{equation}
Here, $\Phi$ is the electrostatic potential and $v$ is the speed.
The mean flow of this Maxwellian is taken to be zero since, as argued in Refs. \cite{HelanderRotation, SugamaHighFlow}, sonic
flows are not permitted in a general stellarator.
Taking $f_{a1}/f_{a0}\sim\rho_{*a}$, the terms of order $\sim\rho_{*a} (v/L)f_{a0}$ in (19) of Ref. \cite{Hazeltine} are then
\begin{equation}
\vpar \vect{b} \cdot\left(\nabla f_{a1}\right)_{W_a,\mu}
-C_{a}
=-\left(\vma + \vE\right)\cdot\nabla\psi \left(\frac{\partial f_{a0}}{\partial \psi}\right)_{W_a}
+\frac{Z_a e}{m_a c}\vpar\vect{b}\cdot\frac{\partial\vect{A}}{\partial t} \frac{\partial f_{a0}}{\partial W_a}
\label{eq:kinetic1}
\end{equation}
where the radial magnetic drift is
\begin{equation}
\vma\cdot\nabla\psi = \frac{m_a c v_{||}^2}{Z_a e B} \vect{b}\times (\vect{b}\cdot\nabla\vect{b})\cdot\nabla\psi
+ \frac{m_a c v_{\perp}^2}{2 Z_a e B^2} \vect{b}\times \nabla B \cdot\nabla\psi
=\frac{m_a c}{2 Z_a e B^2}\left(v_{||}^2 + \frac{v_{\bot}^2}{2}\right) \vect{b}\times \nabla B \cdot\nabla\psi
\end{equation}
(exactly true for any $\beta$ in a magnetic equilibrium with isotropic pressure) and the $\vect{E}\times\vect{B}$ drift
is $\vE=(c/B^2)\vect{B}\times\nabla\Phi$.
Here, $\vect{b} = \vect{B}/B$ is the unit vector along the magnetic field, 
$v_{||}$ and $v_{\bot}$ denote the components of velocity parallel and perpendicular to $\vect{B}$,
$2\pi\psi$ is the toroidal flux, $\vect{A}$ is the magnetic vector potential, 
and $C_a$ is the collision term for species $a$, linearized about the Maxwellians (\ref{eq:Maxwellian}).
Subscripts on gradients and partial derivatives indicate the quantities held fixed,
and $\mu=v_{\bot}^2/(2B)$ is the magnetic moment.

Unfortunately, (\ref{eq:kinetic1}) does not contain the physics of $\vect{E}\times\vect{B}$ precession,
since the characteristic curves of this equation correspond only to motion along the magnetic field lines.
Consequently, important transport regimes such as the $\sqrt{\nu}$ regime cannot be obtained
using (\ref{eq:kinetic1}).
To retain $\vect{E}\times\vect{B}$ precession, we also keep the term $(\vE+\vma)\cdot\nabla f_{a1}$
in (\ref{eq:kinetic1}), even though according to the formal ordering it should
appear at next order.  A similar step is made in other stellarator neoclassical
calculations \cite{DKES1, DKES2}.
The mathematical reason why this term is important at low collisionality is that it has different symmetry properties than other, possibly larger, terms in (\ref{eq:kinetic1}). For instance, it 
survives if a bounce average is used to annihilate the first term.
(We will not bounce average the kinetic equation here,
but when the collisionality is low, the solution of the full equation becomes
asymptotically close to the solution of the bounce-averaged equation.)

As shown in Appendix C of Ref. \cite{usPedestal}, we may choose the gauge
for the electromagnetic potentials such that 
\begin{equation}
-c^{-1}\vect{b}\cdot\partial\vect{A}/\partial t = \left<E_{||} B\right> B/\left<B^2\right>
\label{eq:gauge}
\end{equation}
on the right-hand side of (\ref{eq:kinetic1}).
Here, angle brackets denote a flux surface average:
\begin{equation}
\label{eq:FSA}
\left\langle \ldots \right\rangle
=\frac{1}{{V}'}
\int_0^{2\pi }d\theta
\int_0^{2\pi }d\zeta
{\frac{(\ldots)}{\vect{B}\cdot \nabla \zeta }}
\end{equation}
where
${V}'=\int_0^{2\pi } d\theta \int_0^{2\pi } d\zeta / \vect{B}\cdot \nabla \zeta $,
$\theta$ and $\zeta$ are poloidal and toroidal magnetic angles satisfying
\begin{equation}
\vect{B} = \nabla \psi \times \nabla \theta + \iota \nabla\zeta\times\nabla\psi,
\end{equation}
$\iota = 1/q$ is the rotational transform, and $q$ is the safety factor.
Thus, (\ref{eq:kinetic1}) becomes
\begin{equation}
\left(\vpar \vect{b} + \vE + \vma\right) \cdot\left(\nabla f_{a1}\right)_{W_a,\mu}
-C_{a}
=-\left(\vma + \vE\right)\cdot\nabla\psi \left(\frac{\partial f_{a0}}{\partial \psi}\right)_{W_a}
+\frac{Z_a e}{T_a}\vpar\frac{B\left<E_{||} B\right>}{\left< B^2\right>} f_{a0}.
\label{eq:kinetic2}
\end{equation}
Even if the radial electric field is considered an input, this form of the kinetic equation remains
nonlinear in the unknowns since the $\nabla f_{a1}$ term depends on the variation of $\Phi$
on a flux surface, and this variation is an unknown like $f_{a1}$.

To make the problem linear, we make use of the fact that the electrostatic potential is nearly a flux function.
We define $\Phi_0 = \left<\Phi\right>$ and $\Phi_1 = \Phi-\Phi_0$.
We assume $\Phi_1 \ll \Phi_0$,
and we will show shortly that this assumption is self-consistent.
Since $e \Phi_0/T_a \sim 1$ in the drift ordering, then $e \Phi_1/T_a \ll 1$.
We do not expand in the ion charge $Z_a$.
Equation (\ref{eq:Maxwellian}) then gives
$f_{a0} \approx F_{a}\left[ 1 - Z_a e \Phi_1/T_a\right]$ where
\begin{equation}
F_{a}=n_a(\psi) \left[\frac{m_a}{2\pi T_a(\psi)}\right]^{3/2}\exp\left(-\frac{m_a v^2}{2 T_a(\psi)}\right)
\end{equation}
and $n_a = \eta_a \exp(-Z_a e \Phi_0/T_a)$ is the leading order density.
We define the leading-order total energy
$W_{a0} =v^2/2+ Z_a e\Phi_0 /m_a$,
and leading-order $\vect{E}\times\vect{B}$ drift
$\vEo=(c/B^2)(d\Phi_0/d\psi)\vect{B}\times\nabla\psi$.
As the relative differences between $f_{a0}$ and $F_a$, between $W_a$ and $W_{a0}$, and between $\vE$ and $\vEo$
are all small, we may replace the former quantities with the latter ones in (\ref{eq:kinetic2}).
At the same time, we note
\begin{equation}
\frac{\vE\cdot\nabla\psi}{\vma\cdot\nabla\psi} \sim \frac{1}{\epsilon} \frac{Z_a e \Phi_1}{T_a}
\label{eq:radialExB}
\end{equation}
where $\epsilon$ is the relative variation of $B$ on a flux surface, and taking the ratio (\ref{eq:radialExB}) to be small, the $\vE\cdot\nabla\psi$
term in (\ref{eq:kinetic2}) may be neglected. Thus, we obtain
\begin{equation}
\left(\vpar \vect{b} + \vEo + \vma\right) \cdot\left(\nabla f_{a1}\right)_{W_{a0},\mu}
-C_{a}
=-(\vma\cdot\nabla\psi) \left(\frac{\partial F_{a}}{\partial \psi}\right)_{W_{a0}}
+\frac{Z_a e}{T_a}\vpar\frac{B\left<E_{||} B\right>}{\left< B^2\right>} F_{a},
\label{eq:kinetic3}
\end{equation}
where $C_a$ is now the collision operator linearized about $F_a$ rather than $f_{a0}$,
\begin{equation}
\left(\frac{\partial F_{a}}{\partial \psi}\right)_{W_{a0}}
=\left[ \frac{1}{p_a}\frac{dp_a}{d\psi} + \frac{Z_a e}{T_a} \frac{d\Phi_0}{d\psi} + \left( x_a^2-\frac{5}{2}\right) \frac{1}{T_a} \frac{dT_a}{d\psi}\right]F_a,
\end{equation}
and $x_a = v/v_a$.
If $F_a$ and $\Phi_0$ are considered known, then (\ref{eq:kinetic3}) is now linear in the unknowns $f_{a1}$,
and $\Phi_1$ has decoupled from the kinetic equations.

We note that in some circumstances the ratio (\ref{eq:radialExB}) may not be small \cite{HoKulsrud},
particularly for impurities \cite{JOSE} with $Z_a \gg 1$.
However, treating the ratio (\ref{eq:radialExB}) as finite leads to a kinetic equation
that is nonlinear in the unknowns.  We neglect these nonlinear effects of $\Phi_1$ in the present
linear study, but such effects will be important to examine in future work.

For numerical computations, it is convenient to use coordinates for which the ranges of allowed values
are independent of the other coordinates.  As $W_{a0}$ and $\mu$ do not have this property, it is convenient
to switch to coordinates $x_a$ and $\xi = \vpar/v$.  Carrying out this change of variables on the first term
of (\ref{eq:kinetic3}), we find
\begin{equation}
\dot{\vect{r}} \cdot\left(\nabla f_{a1}\right)_{W_{a0},\mu}
=
\dot{\vect{r}}\cdot \left(\nabla f_{a1}\right)_{x_a,\xi}
+\dot{x}_a \left(\frac{\partial f_{a1}}{\partial x_a}\right)_{\vect{r},\xi}
+\dot{\xi}_a \left(\frac{\partial f_{a1}}{\partial \xi}\right)_{\vect{r},x_a},
\end{equation}
where $\vect{r}$ denotes the position vector,
\begin{equation}
\dot{\vect{r}} = \vpar \vect{b} + \vEo + \vma,
\label{eq:rdot1}
\end{equation}
\begin{equation}
\dot{x}_a = (\vma\cdot\nabla\psi) \left(-\frac{x_a}{2T_a}\frac{dT_a}{d\psi} - \frac{Z_a e}{2T_a x_a}\frac{d\Phi_0}{d\psi}\right),
\label{eq:xdot1}
\end{equation}
and
\begin{equation}
\dot{\xi}_a = -\frac{1-\xi^2}{2 B \xi} \vpar \vect{b}\cdot\nabla B
+\xi(1-\xi^2) \frac{c}{2 B^3} \frac{d\Phi_0}{d\psi}\vect{B}\times\nabla\psi\cdot\nabla B
-\frac{1-\xi^2}{2 B \xi} \vma\cdot\nabla B
.
\label{eq:xidot1}
\end{equation}
For the rest of this work, we will neglect the $\vma$ term in (\ref{eq:rdot1}), the $dT_a/d\psi$ term
in (\ref{eq:xdot1}), and the $\vma\cdot\nabla B$ term in (\ref{eq:xidot1}),
for several reasons.
First, if the  $\vma$ term in (\ref{eq:rdot1}) was retained, we would need to solve a 5D rather
than 4D problem due to the radial coupling (i.e. $\psi$ appearing
as a derivative rather than merely as a parameter). Second,
once radial coupling is dropped, we must also drop the $dT_a/d\psi$ term
in (\ref{eq:xdot1}) and the $\vma\cdot\nabla B$ term in (\ref{eq:xidot1})
in order for $\mu$ to be conserved.
Third, dropping these terms conveniently eliminates all dependence of the transport matrix (defined in section \ref{sec:transportMatrix})
on $d T_a/d\psi$, $dB/d\psi$,
and $\rho_*$.
Fourth, dropping these terms amounts to taking the limit $\rho_* \to 0$
(while keeping the $d\Phi_0/d\psi$ terms finite), 
and this limit is already complicated and interesting to explore without the extra complexity
of finite-$\rho_*$ corrections.
Fifth, 
we wish to focus on the effects of the radial electric field.
The omitted terms may be important in other situations, but here our primary interest is the treatment of the $d\Phi_0/d\psi$ terms.
Finally,
these omitted terms would significantly complicate the analysis in section \ref{sec:sources},
in which we will examine moments of the kinetic equation.

Our kinetic equation then becomes
\begin{equation}
\dot{\vect{r}}\cdot \left(\nabla f_{a1}\right)_{x_a,\xi}
+\dot{x}_a \left(\frac{\partial f_{a1}}{\partial x_a}\right)_{\vect{r},\xi}
+\dot{\xi}_a \left(\frac{\partial f_{a1}}{\partial \xi}\right)_{\vect{r},x_a}
-C_{a}
=-(\vma\cdot\nabla\psi) \left(\frac{\partial F_{a}}{\partial \psi}\right)_{W_{a0}}
+\frac{Z_a e}{T_a}\vpar\frac{B\left<E_{||} B\right>}{\left< B^2\right>} F_{a},
\label{eq:kinetic4}
\end{equation}
where the effective particle trajectory equations are
\begin{eqnarray}
\label{eq:fullTrajectories}
\dot{\vect{r}} &=& \vpar \vect{b} + \frac{c}{B^2}\frac{d\Phi_0}{d\psi} \vect{B}\times\nabla\psi,
\\
\dot{x}_a &=& -(\vma\cdot\nabla\psi) \frac{Z_a e}{2T_a x_a}\frac{d\Phi_0}{d\psi},
\nonumber \\
\dot{\xi}_a &=& -\frac{1-\xi^2}{2 B \xi} \vpar \vect{b}\cdot\nabla B
+\xi(1-\xi^2) \frac{c}{2 B^3} \frac{d\Phi_0}{d\psi}\vect{B}\times\nabla\psi\cdot\nabla B
.\nonumber
\end{eqnarray}
We will refer to (\ref{eq:fullTrajectories}) as the ``full trajectories.''

The $d\Phi_0/d\psi$ terms in $\dot{x}_a$ and $\dot{\xi}_a$ may be interpreted as a finite orbit width effect.
As a particle drifts radially, it experiences a varying electrostatic potential (even if the potential is a flux function.)
Thus the potential energy of the particle changes, so to maintain a constant total energy, the kinetic energy must change at
an equal and opposite rate, giving rise to the $d\Phi/d\psi$ term in $\dot{x}_a$.
Then to conserve $\mu$ while $v$ changes, $\xi$ must also change appropriately,
giving rise to the $d\Phi_0/d\psi$ term in $\dot{\xi}_a$.  Without these $d\Phi_0/d\psi$
terms in $\dot{x}_a$ and $\dot{\xi}_a$, $\mu$ will not be conserved,
whereas you can verify that $\mu$ is indeed conserved by (\ref{eq:fullTrajectories}).
Note that the $d\Phi_0/d\psi$ term in $\dot{\vect{r}}$
is the same order in the $\rho_*$ expansion as the
$d\Phi_0/d\psi$ terms in $\dot{x}_a$ and $\dot{\xi}_a$,
suggesting that if the former term is retained, the latter terms should be retained as well.

A large number of stellarator neoclassical codes \cite{DKES1, DKES2, BeidlerBigBenchmarking} effectively
solve (\ref{eq:kinetic4}) with the alternative trajectory equations
\begin{eqnarray}
\label{eq:DKESTrajectories}
\dot{\vect{r}} &=& \vpar \vect{b} + \frac{c}{\left<B^2\right>}\frac{d\Phi_0}{d\psi} \vect{B}\times\nabla\psi,
\\
\dot{x}_a &=& 0,
\nonumber \\
\dot{\xi}_a &=& -\frac{1-\xi^2}{2 B \xi} \vpar \vect{b}\cdot\nabla B
.\nonumber
\end{eqnarray}
We refer to these equations as the ``DKES trajectories,'' in light of their use in the widely applied code DKES \cite{DKES1, DKES2}.
These trajectories differ from (\ref{eq:fullTrajectories}) both in the neglect of the $d\Phi_0/d\psi$ terms in $\dot{x}_a$ and $\dot{\xi}_a$,
and in the replacement $B^2 \to \left< B^2 \right>$ in $\dot{\vect{r}}$.
The motivation for approximating the $\vect{E}\times\vect{B}$ drift in this matter will be clarified in section \ref{sec:sources}.
As shown in Refs. \cite{meMonoenergetic, finiteErOmnigenous}, in a symmetric magnetic field, the model (\ref{eq:DKESTrajectories}) possesses a conserved quantity
which is equal to $\mu$ when $d\Phi_0/d\psi=0$ but which differs from $\mu$ when $d\Phi_0/d\psi\neq 0$.

For comparison, we will also consider the following set of trajectory equations:
\begin{eqnarray}
\label{eq:partialTrajectories}
\dot{\vect{r}} &=& \vpar \vect{b} + \frac{c}{B^2}\frac{d\Phi_0}{d\psi} \vect{B}\times\nabla\psi,
\\
\dot{x}_a &=& 0,
\nonumber \\
\dot{\xi}_a &=& -\frac{1-\xi^2}{2 B \xi} \vpar \vect{b}\cdot\nabla B
\nonumber
\end{eqnarray}
which will be referred to as the ``partial trajectories.''  Equations (\ref{eq:partialTrajectories})
represent an intermediate step between (\ref{eq:DKESTrajectories}) and (\ref{eq:fullTrajectories}),
in that (\ref{eq:partialTrajectories}) includes the correct $\vect{E}\times\vect{B}$ drift, but not the
$d\Phi_0/d\psi$ terms in $\dot{x}_a$ and $\dot{\xi}_a$ required to conserve $\mu$.

Note that for both the DKES and full trajectories, the left-hand side of the
kinetic equation (\ref{eq:kinetic4})
can be written in the conservative form
\begin{equation}
\frac{1}{J}\left[
\nabla\cdot\left( J \dot{\vect{r}}_a f_{a1} \right) 
+ \frac{\partial}{\partial \xi}\left( J \dot{\xi}_a f_{a1} \right)
+ \frac{\partial}{\partial x_a} \left( J \dot{x}_a f_{a1} \right)
\right] - C_a
\label{eq:conservativeForm}
\end{equation}
where $J=x_a^2$ is the Jacobian of the transformation between Cartesian velocity coordinates
and the coordinates $x_a$, $\xi$, and gyrophase. However, for the partial trajectories,
the left-hand side of (\ref{eq:kinetic4}) is not equivalent
to (\ref{eq:conservativeForm}).

For all three trajectory models, the quasineutrality equation is effectively decoupled from the kinetic equation (\ref{eq:kinetic4}).
At leading order, quasineutrality implies $\sum_a Z_a n_a = 0$.
At next order, noting that both $f_{a0}$ and $f_{a1}$ contribute to density variation on a flux surface,
\begin{equation}
\sum_a \left( -\frac{Z_a^2 e \Phi_1}{T_a}n_a + Z_a \int d^3v\, f_{a1} \right) = 0.
\label{eq:quasineutrality}
\end{equation}
This equation may be solved for $\Phi_1$, giving the variation of the potential on a flux surface.
It follows that $e \Phi_1 / T_a \sim f_{a1} / f_{a0} \sim \rho_{*a}$, so our earlier assumption that
$e \Phi_1 / T_a \ll 1$ is self-consistent.

Several choices can be made for the collision operator.
The most accurate linear option is the Fokker-Planck operator \cite{RMJ,PerBook}
linearized about the Maxwellians: $C_a = \sum_b C_{ab}^{\ell}$,
where $C_{ab}^{\ell} = C_{ab}\{f_{a1},F_{b}\} + C_{ab} \{ F_{a}, f_{b1}\}$
and $C_{ab}$ is the full bilinear Fokker-Planck operator between species $a$ and $b$.
This linearized operator may be written in many forms, and for numerical implementation,
we find it convenient to use the form detailed in equations (14)-(16) of Ref. \cite{speedGrids}.

A simpler option used in many codes is the pitch-angle scattering operator \cite{BeidlerBigBenchmarking}.
This operator lacks several properties of the linearized Fokker-Planck operator,
such as the momentum conservation property $\int d^3v \, v_{||} C_{aa}^{\ell}=0$.
Several more accurate approximate operators have been used in the literature.
One such operator we will consider later consists of the pitch-angle scattering operator
plus an ad-hoc momentum-restoring term, given for the case of self-collisions
by eq (3.69) in Ref. \cite{PerBook}.

\section{Particle and energy moment equations, conservation properties, and sources}
\label{sec:sources}

If one attempts to solve the kinetic equation (\ref{eq:kinetic4}) numerically using either the full or partial trajectories
and $E_r \neq 0$, unphysical results will be obtained, with
the numerical solution not converging as resolution parameters are increased.
We now explore the reason for this behavior.
We will then describe a modified form of the kinetic equation which robustly produces more sensible results.
The issues discussed in this section are related to moment equations for mass and energy; momentum has a different status
and will be examined in the next section.

Consider the result of applying the operation
\begin{equation}
\left< \int d^3v(\ldots)\right>
\label{eq:massConservation}
\end{equation}
to the kinetic equation (\ref{eq:kinetic4}) for each of the trajectory models (\ref{eq:fullTrajectories})-(\ref{eq:partialTrajectories}).
This operation annihilates the streaming and mirror terms, the collision operator, and the
inhomogeneous drive terms.
The operation (\ref{eq:massConservation}) effectively produces a flux-surface-averaged
mass conservation equation for each model.  For the full trajectories and DKES trajectories,
the $d\Phi_0/d\psi$ terms are also annihilated by
(\ref{eq:massConservation}), so the
resulting mass conservation equation is just $0=0$. However, for the partial trajectories, the $d\Phi_0/d\psi$
term ($\vEo\cdot\nabla f_{a1}$) is not annihilated by (\ref{eq:massConservation}),
leaving
\begin{equation}
c \left< \frac{1}{B^2}
\vect{B}\times\nabla\psi\cdot\nabla
\int d^3v\, f_{a1}\right> \frac{d\Phi_0}{d\psi}=0.
\label{eq:unphysicalMassConservation}
\end{equation}
Thus, a nonzero $d\Phi_0/d\psi$ gives a singular perturbation to the $d\Phi_0/d\psi=0$ limit in this partial trajectory model:
the $d\Phi_0/d\psi=0$ solution for $f_{a1}$ need not satisfy 
$\left< (1/B^2) \vect{B}\times\nabla\psi\cdot\nabla\int d^3v\, f_{a1}\right>=0$,
so $f_{a1}$ must change dramatically as $E_r$ is raised from 0 to a small nonzero value,
a behavior which is unphysical.
When $d\Phi_0/d\psi \not= 0$, (\ref{eq:unphysicalMassConservation}) constrains $f_{a1}$ in an unphysical
manner, for there is no analogue to (\ref{eq:unphysicalMassConservation}) in the more accurate
averaged fluid mass conservation equation $0=\left< \partial N_a/\partial t + \nabla\cdot(N_a \vect{V}_a)\right>$
(i.e. the moment of the full Fokker-Planck equation with no expansion in $\rho_*$ or other parameters),
where $N_a$ and $\vect{V}_a$ are the full fluid density and velocity.
The unphysical nature of (\ref{eq:unphysicalMassConservation}) can also be seen from the fact that when the
$d\Phi_0/d\psi$ terms in $\dot{x}_a$ and $\dot{\xi}_a$ are retained in the more accurate trajectories (\ref{eq:fullTrajectories}),
these terms precisely cancel (\ref{eq:unphysicalMassConservation}).

Similarly, we can obtain an averaged energy conservation equation for each trajectory model by applying the operation
\begin{equation}
\sum_a \left< \int d^3v\frac{m_a v^2}{2}(\ldots)\right>
\end{equation}
to (\ref{eq:kinetic4}). Again, the result is $0=0$ for the DKES trajectories. However,
this time both the full and partial trajectory models give nonzero results: the partial trajectories give
\begin{equation}
c \sum_a\left< \frac{1}{B^2}
\vect{B}\times\nabla\psi\cdot\nabla
\int d^3v \frac{m_a v^2}{2} f_{a1}\right> \frac{d\Phi_0}{d\psi}=0.
\label{eq:unphysicalEnergyConservation1}
\end{equation}
and the full trajectories give
\begin{equation}
-c \sum_a\left< \frac{1}{B^2}
\vect{B}\times\nabla\psi\cdot\nabla
\int d^3v \frac{m_a v^2}{2}\frac{(1+\xi^2)}{2} f_{a1}\right> \frac{d\Phi_0}{d\psi}=0.
\label{eq:unphysicalEnergyConservation2}
\end{equation}
The quantity multiplying $d\Phi_0/d\psi$ in (\ref{eq:unphysicalEnergyConservation2})
is proportional to the radial current
$\sum_a Z_a \left< \int d^3v\, f_{a1} \vma\cdot\nabla\psi\right>$,
so it vanishes naturally when $E_r$ is at the ambipolar value.  However, as
the radial current would usually not be zero when $E_r=0$, (\ref{eq:unphysicalEnergyConservation2})
again implies a small nonzero $E_r$ would be a singular perturbation of the
$E_r=0$ limit.

One motivation for use of the DKES trajectory model is now apparent:
it is the only model (of the three condered here) that avoids the imposition of one or more unphysical
constraints on the distribution function when $d\Phi_0/d\psi \ne 0$,
constraints which cause an $E_r \ne 0$ calculation to be a singular perturbation
of an $E_r = 0$ calculation.

The aforementioned problems with the partial and full trajectory models may be eliminated 
in the following manner.
The kinetic equation becomes well behaved if we introduce particle and heat sources
\begin{equation}
S_a(\psi, v) = \Sap(\psi) F_{a}(\psi,v) \left[ x_a^2 - \frac{5}{2}\right] + \Sah(\psi) F_a(\psi, v) \left[x_a^2 - \frac{3}{2}\right]
\label{eq:sources}
\end{equation}
where $\Sap$ and $\Sah$ are considered to be unknowns.
(The factors involving $x_a^2$ in (\ref{eq:sources}) are chosen so $\Sap$ provides a particle source but no heat source, while $\Sah$ provides
a heat source but no particle source.)
As these two new unknowns are now included in the system of equations on each flux surface, we must supply an equal number of additional constraints.
The constraints we supply are $\left< \int d^3v\, f_{a1}\right>=0$ and $\left< \int d^3v\, v^2 f_{a1}\right>=0$,
the sensible requirements that all the flux-surface-averaged density and pressure reside in $F_a$ rather than $f_{a1}$.
When $S_a$ is included in the kinetic equation, new terms proportional to $\Sap$ and/or $\Sah$
now appear in the mass and energy conservation equations such as (\ref{eq:unphysicalMassConservation})-(\ref{eq:unphysicalEnergyConservation2}).
These conservation equations imply that when $d\Phi_0/d\psi=0$, $\Sap$ and $\Sah$ must vanish.
However, now when $d\Phi_0/d\psi$ is increased from 0 to a small finite number, the sources can turn on
to satisfy  (\ref{eq:unphysicalMassConservation})-(\ref{eq:unphysicalEnergyConservation2}),
eliminating the singular perturbation in $f_{a1}$.  We find that numerical results are then well behaved,
converging appropriately as numerical resolution parameters are increased, and smoothly going to the
$E_r=0$ results as $E_r$ is decreased.

We do not claim that the method proposed here is an ideal solution: the sources (\ref{eq:sources})
are ad-hoc and are not derived rigorously. However, by the techniques proposed here, we can at least
compare the three different trajectory models, and for most experimentally relevant values of $E_r$, we will show
that the three models give nearly identical results. 
And as already mentioned, the source terms for the full trajectory model are both zero when the radial electric field equals the value required 
for ambipolarity, so for this model the source terms are really a numerical expedient that
do not affect the transport computations in the end.

This system of sources and constraints solves not only the problem described above when $E_r \ne 0$,
but also a different problem that remains even when $E_r=0$ and/or when the DKES trajectories are used:
the kinetic equation has a null space.
If the conditions $\left< \int d^3v \,f_{a1}\right>=0$ and $\left< \int d^3v\, v^2 f_{a1}\right>=0$
were not imposed, any linear combination of $F_{a}$ and $F_{a} v^2$ could be added to one solution
of the kinetic equation to obtain another solution. Upon discretization, one would obtain a non-invertible
(or at least very poorly conditioned) linear system, but the imposition of these two extra constraints
makes the system of equations invertible. 

Such is the case when the full linearized Fokker-Planck 
collision operator is used, but the situation is different when either the pitch-angle scattering
operator or momentum-conserving model operator are used instead, for then the kinetic equation has a larger null space:
any function of $v$ is then a homogeneous solution of the kinetic equation.
As the dimension of the null space is then equal to $N_x$ (the number of grid points in $x_a$) rather than
2, it takes $N_x$ rather than 2 constraint equations to eliminate the null space for these collision operators.
We choose these $N_x$ constraints to be $\left< \int_{-1}^1 d\xi\, f_{a1}\right>=0$ (imposed at each grid point in $x_a$.)
To keep the linear system square, we must then have $N_x$ rather than 2 unknowns related to the sources. 
This is accomplished by letting the source be a general function of $x_a$ instead
of (\ref{eq:sources}) when either the pitch-angle scattering or momentum-conserving model collision operator are used.
This alternative system of $N_x$ sources and constraints is an equally reasonable solution to the earlier
conservation problem.

To summarize, the sources and extra constraint equations serve two independent purposes.
First, when $E_r \ne 0$, the sources are needed to eliminate the conservation problems, and the extra constraints then keep the linear system square (number of equations = number of unknowns) upon discretization.
Second, even when $E_r = 0$, and even for the DKES trajectories in which sources are not required, the constraints are needed to eliminate the null space in the kinetic equation, and the source terms are a convenient way to keep the linear system square upon discretization.
The first problem can be solved with either the source (\ref{eq:sources}) or a general speed-dependent source $S_a(x_a)$.
However, to solve the second problem, the number of constraints should match the dimensionality of the null space.
For this reason, we apply the source (\ref{eq:sources}) with 2 constraints when the Fokker-Planck operator is used,
while we apply the general speed-dependent source $S_a(x_a)$ with $N_x$ constraints when either of the other two collision operators is used.

\section{Momentum moment equations}
\label{sec:momentum}

Parallel momentum has a different status to density and energy, in that density and energy are conserved by
the collisionless motion while parallel momentum is not, due to the mirror force.
(For example, considering the case of a single ion species with $d\Phi_0/d\psi=0$, $F_i$ and $v^2 F_i$ are homogeneous
solutions to the kinetic equation, whereas $v_{||} F_i$ is not.)
A consequence is that there does not appear to be a false constraint for $E_r \ne 0$ arising
from the $\left< \int d^3 v\, v_{||} (\ldots)\right>$ moment of the various forms of the kinetic equation,
i.e. there is no analogue to (\ref{eq:unphysicalMassConservation}), (\ref{eq:unphysicalEnergyConservation1}),
or (\ref{eq:unphysicalEnergyConservation2}) for momentum.
When the momentum moment of the various forms of the drift-kinetic equation is taken,
even if a factor of $B$ or $1/B$ is included in the flux surface average, 
a collisionless term remains that is not proportional to $d\Phi_0/d\psi$.
Consequently, for all the trajectory models,
$d\Phi_0/d\psi=0$ is a well behaved rather than singular
limit of the momentum moment equation.

Nonetheless, it is interesting to compare the $\int d^3v\, m_a v_{||} (\ldots)$ moment equations
for each drift-kinetic trajectory model to the full parallel momentum fluid equation. This later equation,
the moment of the full Fokker-Planck equation, is
\begin{equation}
0 = -\vect{b}\cdot\left( \nabla\cdot\tens{\Pi}_a\right)
+ Z_a e n_a E_{||} + F_{a||}
\label{eq:fluidParallelMomentum}
\end{equation}
where $\tens{\Pi_a} = m_a \int d^3v\, f_a \vect{v}\vect{v}$ is the stress tensor
and $F_{a||}$ is the parallel component of friction.
First, consider the case of no radial electric field.
Recalling
$f_a = F_a(\psi)\left[ 1 - Z_a e \Phi_1/T_a\right] + f_{a1}$, 
the stress tensor is given to the accuracy needed by 
$\tens{\Pi}_a \approx p_a(\psi)\left[1 - Z_a e \Phi_1/T_a\right] \tens{I} + \tens{\Pi}_{a1}$
where $\tens{\Pi}_{a1} = p_{a 1 \bot} \tens{I} + (p_{a1||} - p_{a1\bot}) \vect{b}\vect{b}$,
$p_{a1||} = m_a \int d^3v\, f_{a1} v_{||}^2$, and $p_{a1\bot} = m_a \int d^3v\, f_{a1} v_{\bot}^2/2$.
Notice the $\int d^3v\, m_a v_{||} (\ldots)$ moment of the streaming and mirror terms in (\ref{eq:kinetic4})-(\ref{eq:fullTrajectories}) is
\begin{eqnarray}
\int d^3v\, m_a v_{||} \left[ v_{||} \vect{b}\cdot\nabla f_{a1} - \frac{1-\xi^2}{2B}v(\vect{b}\cdot\nabla B) \frac{\partial f_{a1}}{\partial \xi} \right]
& = &
\vect{b}\cdot\nabla p_{a1||} + \frac{ p_{a1\bot} - p_{a1 ||}}{B} \vect{b}\cdot\nabla B \nonumber \\
& = &
\vect{b} \cdot \left(\nabla \cdot \tens{\Pi}_{a1} \right).
\end{eqnarray}
Using this result with (\ref{eq:gauge}), the $m_a v_{||}$ moment of 
the drift-kinetic equation (\ref{eq:kinetic4})-(\ref{eq:fullTrajectories}) matches
the full fluid parallel momentum equation (\ref{eq:fluidParallelMomentum})
at least when $E_r=0$.

Now consider how the situation changes when a radial electric field is introduced. We first
compute the change to the fluid parallel momentum equation caused by a new contribution
to the viscosity.
Examining the $\left( \vect{E} + c^{-1} \vect{v}\times\vect{B}\right) \cdot \nabla_{\vect{v}} f_a$ terms in 
the full Fokker-Planck equation and integrating in gyrophase, one sees the gyrophase-dependent part of the 
distribution function $\tilde{f}_a$ will include the following terms proportional to the electric field:
\begin{equation}
\tilde{f}_{aE} = \frac{c}{B} \vect{v}\cdot\vect{b}\times\vect{E} \left[ \frac{1}{v} \frac{\partial \bar{f}_a}{\partial v}
-\frac{\xi}{v^2} \frac{\partial \bar{f}_a}{\partial \xi} \right]
\label{eq:fTilde}
\end{equation}
as reflected (using different independent variables) in eq (17) of Ref. \cite{Hazeltine} and eq (6) of Ref. \cite{SimakovCattoSecondOrderDKE}.
Here, $\bar{f}_a$ is the gyrophase-independent part of the distribution function.
The associated contribution $\tens{\Pi}_{aE} = m_a \int d^3v\, \tilde{f}_{aE}\vect{v}\vect{v}$
to the pressure tensor is calculated in equations (27)-(36) of Ref. \cite{SimakovCattoSecondOrderDKE},
with the result
\begin{equation}
\tens{\Pi}_{aE} = m_a n_a V_{a||} \left( \vect{b}\vect{v}_E + \vect{v}_E \vect{b} \right)
\label{eq:PiE}
\end{equation}
where $n_a V_{a||} = \int d^3v\, f_a v_{||}$. Note that this contribution to the stress tensor is a part of the gyroviscosity and is off-diagonal.
Using $\vect{E} \approx -\nabla \Phi_0(\psi)$ and $\vect{B} \cdot \left\{ \nabla \cdot \left[ \vect{B}\vect{B}\times\nabla\psi + (\vect{B}\times\nabla\psi) \vect{B} \right] \right\}
=2B^2\nabla\psi\cdot\nabla\times\vect{B}=0$,
we then find the contribution to the parallel momentum equation (\ref{eq:fluidParallelMomentum}) from (\ref{eq:fTilde})-(\ref{eq:PiE}) is
\begin{equation}
\vect{b}\cdot(\nabla\cdot\tens{\Pi}_{aE}) = 
c m_a B (d\Phi_0/d\psi)
\vect{B}\times\nabla\psi\cdot\nabla\left(n_a V_{a||}/B^3\right).
\label{eq:PiEParallel}
\end{equation}

For comparison, let us consider the $\int d^3v\, m_a v_{||} (\ldots)$ moment of the radial electric field
terms in the drift-kinetic equation for various trajectory models, to see if the results agree with (\ref{eq:PiEParallel}).
For the full trajectories, the moment of the $d\Phi_0/d\psi$ terms in (\ref{eq:kinetic4})-(\ref{eq:fullTrajectories}) is
\begin{eqnarray}
\int d^3v\, m_a v_{||} \left[ 
\frac{c}{B^2} \frac{d\Phi_0}{d\psi} \vect{B}\times\nabla\psi\cdot\nabla f_{a1}
- (\vma \cdot\nabla\psi) \frac{Z_a e}{2 T_a x_a} \frac{d\Phi_0}{d\psi} \frac{\partial f_{a1}}{\partial x_a}
\right.
\nonumber \\
\left.
+ \xi(1-\xi^2) \frac{c}{2B^3} \frac{d\Phi_0}{d\psi} (\vect{B}\times\nabla\psi\cdot\nabla B)\frac{\partial f_{a}}{\partial \xi}
\right]
\nonumber \\
= 
c m_a B (d\Phi_0/d\psi)
\vect{B}\times\nabla\psi\cdot\nabla\left(n_a V_{a||}/B^3\right),
\label{eq:fullViscosity}
\end{eqnarray}
obtained by integrating by parts in $x_a$ and $\xi$.
Thus, the full-trajectory model agrees with the full fluid parallel momentum equation: (\ref{eq:fullViscosity}) = (\ref{eq:PiEParallel}).
However, this agreement is
not shared by the DKES model: the moment of the $d\Phi_0/d\psi$ term in (\ref{eq:DKESTrajectories}) is
\begin{eqnarray}
\int d^3v\, m_a v_{||} \left[ 
\frac{c}{\left< B^2 \right> } \frac{d\Phi_0}{d\psi} \vect{B}\times\nabla\psi\cdot\nabla f_{a1}
\right]
= 
\frac{c m_a}{ \left< B^2 \right>}
\frac{d\Phi_0}{d\psi}
\vect{B}\times\nabla\psi\cdot\nabla (n_a V_{a||}),
\label{eq:DKESViscosity}
\end{eqnarray}
which does not equal (\ref{eq:PiEParallel}). The corresponding result for the partial trajectories,
obtained by replacing $\left< B^2 \right> \to B^2 $ in (\ref{eq:DKESViscosity}), also does not match
(\ref{eq:PiEParallel}). Thus, the DKES and partial trajectory models do not correctly account for the parallel
viscous force as the full trajectory model does.

We close this section by noting another important difference between the trajectory models
related to the parallel momentum equations.
Consider the case of \emph{quasisymmetry}, which is the condition 
that
$\vect{B}\times\nabla\psi\cdot\nabla B = A(\psi) \vect{B}\cdot\nabla B$ for some flux function
$A(\psi)$ \cite{PerAndreiPRL}.  It was known previously \cite{PerAndreiPRL} that when the $E_r$ terms are not included
in the trajectories (but retained in the $\partial F_a/\partial \psi$ drive term in (\ref{eq:kinetic4})), the radial neoclassical
current vanishes for all values of $d \Phi_0/d\psi$
if and only if the flux surface is quasisymmetric. This property of quasisymmetric flux surfaces
is known as intrinsic ambipolarity.
Here, we show that intrinsic ambipolarity persists in quasisymmetric geometry
when the $E_r$ terms are retained in the full trajectory
drift-kinetic equation, but not for the DKES or partial trajectory kinetic equations.
This result follows from the $-\sum_a Z_a \left< \int d^3v\, A v_{||} / \Omega_a (\ldots)\right>$
moment of the kinetic equations, i.e. a spatially weighted average of the parallel momentum
moment. For the full trajectories, 
(\ref{eq:fullViscosity}) vanishes in this spatial average, leaving
\begin{equation}
\sum_a Z_a \left< \int d^3v\, f_{a1} \vma\cdot\nabla\psi\right>=0,
\label{eq:noCurrent}
\end{equation}
meaning there is no radial current. However, for the DKES and partial trajectory models, the
spatial average does not annihilate the $d\Phi_0/d\psi$ term,
leaving an additional term in (\ref{eq:noCurrent}) proportional to $d\Phi_0/d\psi$,
and therefore the radial current is generally nonzero.  
Consequently, the full trajectory model is the only one of the models that preserves
intrinsic ambipolarity in quasisymmetry for $E_r \ne 0$.
Notice that when the full trajectory model is applied in quasisymmetry,
intrinsic ambipolarity means (\ref{eq:unphysicalEnergyConservation2}) is satisfied even when $d\Phi_0/d\psi \ne 0$,
so the net heat source vanishes for any radial electric field.

\section{Numerical implementation}
\label{sec:numerics}

The SFINCS code solves the drift-kinetic equation (\ref{eq:kinetic4})
with (\ref{eq:sources})
for any of the three trajectory models (\ref{eq:fullTrajectories})-(\ref{eq:partialTrajectories}),
for general nonaxisymmetric nested flux surface geometry, and for an arbitrary number of species.
SFINCS is based on the Fokker-Planck code described in Ref. \cite{speedGrids},
generalized to allow nonaxisymmetry.
SFINCS is also closely related to the radially global Fokker-Planck code for tokamaks described
in Ref. \cite{PERFECT}.
Briefly, the kinetic equation is discretized using finite differences with a 5-point stencil in $\theta$
and $\zeta$, using a truncated Legendre modal expansion in $\xi$, and using
a spectral collocation method in $x_a$.
The time-independent kinetic equation is solved directly (by solving a single sparse linear system),
so the rate of convergence is not limited by the timescale of physical relaxation.
The modifications compared to the code of Ref. \cite{speedGrids} are the following.
(1) $f_{a1}$, $B$, and other geometric operators are allowed to depend on the toroidal angle $\zeta$,
and the numerical grid is expanded to include this new coordinate.
(2) The additional $d\Phi_0/d\psi$ terms in $\vect{\dot{r}}_a$, $\dot{x}_a$, and $\dot{\xi}_a$
are included.
(3) The additional collision operators discussed above are included.
(4) The extra constraint equations and sources are implemented as in (19) of Ref. \cite{PERFECT}.
Specifically, considering first the case of a single species for simplicity, the linear system has the block structure
\begin{equation}
\begin{array}{r}
\mbox{Kinetic equation}\;\;\{ \\
\left<\int d^3v f_{a1}\right>=0 \;\;\{ \\
\left<\int d^3v\; f_{a1} v^2\right>=0\;\; \{
\end{array}
\left(\begin{array}{ccc}
M_{11} & M_{12} & M_{13} \\
M_{21} & 0 & 0 \\
M_{31} & 0 & 0 \end{array}
\right)
\underbrace{
\left(
\begin{array}{c}
f_{a1} \\
\Sap \\
\Sah \end{array}
\right)}_{\mbox{Vector of unknowns}}
 = \left(
\begin{array}{c}
R
 \\
0 \\
0 \end{array}
\right),
\label{eq:blocks}
\end{equation}
where $R$ is the inhomogeneous term (i.e. the right-hand side) from (\ref{eq:kinetic4}),
and the $M$ operators are as follows: $M_{11}$ represents the operator on the left-hand side of (\ref{eq:kinetic4}),
$M_{12}$ and $M_{13}$ represent the $\Sap$ and $\Sah$ terms
in (\ref{eq:sources}) respectively,
and $M_{21}$ and $M_{31}$ represent the aforementioned extra constraint equations introduced.
For the case of multiple particle species, the linear system consists of blocks of the form (\ref{eq:blocks}) for each
species, with coupling between species only through the collision operators in the $M_{11}$ blocks.

The resulting large sparse linear system is solved using the
PETSc\cite{petsc-web-page, petsc-user-ref} library.
A preconditioned iterative Krylov solver is employed,
either GMRES\cite{GMRES} or BICGStab(l)\cite{BICGstabl}.
An effective preconditioner is typically obtained by dropping all coupling in the $x_a$ coordinate,
either for all Legendre modes in $\xi$, or for all but the first one or two Legendre modes.
The preconditioner is $LU$-factorized directly using the SuperLU-dist\cite{superlu1, superlu2} package.

Note that poloidal and toroidal magnetic drifts could be included in the kinetic equation
without increasing the density of the matrix, i.e. without increasing the computational expense
of the method here.  
We do not expect any fundamental new complications to arise
if poloidal and toroidal magnetic drift terms are retained.
However, to include radial drifts acting on $f_{a1}$,
the number of independent variables would increase from 4 to 5 since different flux surfaces
would couple. This increase in dimensionality would be numerically challenging.

The magnetic geometry is specified in Boozer coordinates $\theta$ and $\zeta$, in which
\begin{equation}
\vect{B} =  \beta(\psi,\theta,\zeta)\nabla\psi + I(\psi)\nabla\theta + G(\psi) \nabla\zeta.
\end{equation}
Here, 
$cI/2$ is the toroidal current inside the flux surface, and $cG/2$ is the poloidal
current outside the flux surface.
The geometric operators needed in the kinetic equation are then
\begin{equation}
\vect{B}\cdot\nabla X = \left( \iota \frac{\partial X}{\partial\theta} + \frac{\partial X}{\partial\zeta}\right)\vect{B}\cdot\nabla\zeta
\end{equation}
and
\begin{equation}
\vect{B}\times\nabla\psi\cdot\nabla X = \left( G \frac{\partial X}{\partial\theta} -I \frac{\partial X}{\partial\zeta}\right)\vect{B}\cdot\nabla\zeta
\end{equation}
where $X$ can be any scalar quantity,
and the inverse coordinate Jacobian is
$\vect{B}\cdot\nabla\zeta = B^2/(G+\iota I)$. Thus, the magnetic geometry
enters the kinetic equation only through the quantities $I$, $G$, $\iota$,
and $B(\theta,\zeta)$.

\section{Ion transport matrix}
\label{sec:transportMatrix}

We will present results of the numerical calculations in terms of the transport matrix $L_{jk}$, defined as follows:
\begin{equation}
\label{eq:matrix}
\left(\begin{array}{c}
\frac{Ze(G+\iota I)}{ncTG} \left< \int d^3v\, f \vm\cdot\nabla\psi\right> \\
\frac{Ze(G+\iota I)}{ncTG} \left< \int d^3v\, f \frac{mv^2}{2T}\vm\cdot\nabla\psi\right> \\
\frac{1}{\vi B_0} \left< B V_{||} \right>
\end{array}\right)
=
\left(\begin{array}{ccc}
L_{11} & L_{12} & L_{13} \\
L_{21} & L_{22} & L_{23} \\
L_{31} & L_{32} & L_{33}
\end{array}\right)
\left(\begin{array}{c}
\frac{GTc}{ZeB_0 v_{i}} \left[ \frac{1}{n}\frac{dn}{d\psi} + \frac{Ze}{T}\frac{d\Phi}{d\psi} -\frac{3}{2T}\frac{dT}{d\psi}\right] \\
\frac{GTc}{Ze B_0 v_{i} T}\frac{dT}{d\psi} \\
\frac{Ze}{T}(G+\iota I)\frac{\left<E_{||} B\right>}{\left<B^2\right>}
\end{array}\right)
\end{equation}
Here, $B_0$ is the $(0,0)$ Fourier mode amplitude of $B(\theta,\zeta)$, and we have dropped $i$ subscripts
where possible to simplify the notation.
When the DKES trajectories (\ref{eq:DKESTrajectories}) are used, it can be shown that $L_{jk}$ is symmetric for any value of $E_r$.
When the trajectories (\ref{eq:fullTrajectories}) or (\ref{eq:partialTrajectories})
are used and $E_r=0$, $L_{jk}$ is symmetric as well. However, when the trajectories (\ref{eq:fullTrajectories}) or (\ref{eq:partialTrajectories}) are used and $E_r \neq 0$,
the transport matrix defined in this manner is generally not symmetric.

Different definitions of the transport matrix have been given elsewhere in the literature \cite{Beidler},
but the definition here has several nice properties. First, the matrix is dimensionless. Second, $L_{jk}$ is symmetric (in the cases described above).
Third, $L_{jk}$ depends on the magnetic geometry and physical parameters only through
$B/B_0$, $I/G$, $\iota$, a normalized collisionality
\begin{equation}
\nu' = \frac{(G+\iota I)\nuii}{\vi B_0},
\label{eq:nuPrime}
\end{equation}
and a normalized electric field
\begin{equation}
E_* = \frac{c G}{\iota \vi B_0}\frac{d\Phi_0}{d\psi},
\label{eq:EStar}
\end{equation}
and not on any other individual parameters such as density, temperature, $G$, etc.
In (\ref{eq:nuPrime}), $\nuii = 4\sqrt{2\pi} n Z^4 e^4 \ln\Lambda/(3 m^{1/2} T^{3/2})$
is the ion-ion collision frequency.
Typically, $I \ll G$ and $G \approx B_0 R$ where $R$ is the major radius of the device,
so $\nu' \approx \nuii R / \vi$.
In axisymmetry, $E_*$ corresponds to the poloidal Mach number: $E_* \approx (B / \Bpol) | \vEo | / \vi$
where $\Bpol$ is the poloidal magnetic field.
Therefore, $E_*$ corresponds to the electric field normalized by the so-called resonant electric field \cite{Beidler}
$\Erres = r\iota \vi B/(R c)$, with $r/R$ the inverse aspect ratio.

Several properties of the matrix $L_{jk}$ are noteworthy. Using the property $\int d^3v (g/F_{\mathrm{i}}) C_{\mathrm{ii}}\{g\} \le 0$
for any $g$, which holds for all three ion-ion collision operators considered here,
then $\sgn(L_{11}) = \sgn(L_{22}) = -\sgn(L_{33}) = -\sgn((G+\iota I)/B_0)$.
This property holds when $E_r=0$, and it holds when $E_r \ne 0$ for the DKES trajectories, but it
may not hold when $E_r \ne 0$ for the partial or full trajectories.
Second, for all three trajectory models, the elements $L_{jk}$
are independent of the sign of the electric field: $L_{jk}(E_*) = L_{jk}(-E_*)$,
assuming the stellarator symmetry property $B(\theta,\zeta) = B(-\theta,-\zeta)$
for some choice of the origin of $\theta$ and $\zeta$.
This symmetry of $L_{jk}$ follows from a symmetry in the kinetic equation:
if the signs of 
$\theta$, $\zeta$, $v_{||}$, and $d\Phi_0/d\psi$ are all reversed
in (\ref{eq:kinetic4}), the sign of $f_{i1}$ will reverse,
leaving the left-hand side of (\ref{eq:matrix}) unchanged.

\section{Comparison of $E_r$ terms}
\label{sec:ErComparison}

\begin{figure}[h!]
\includegraphics{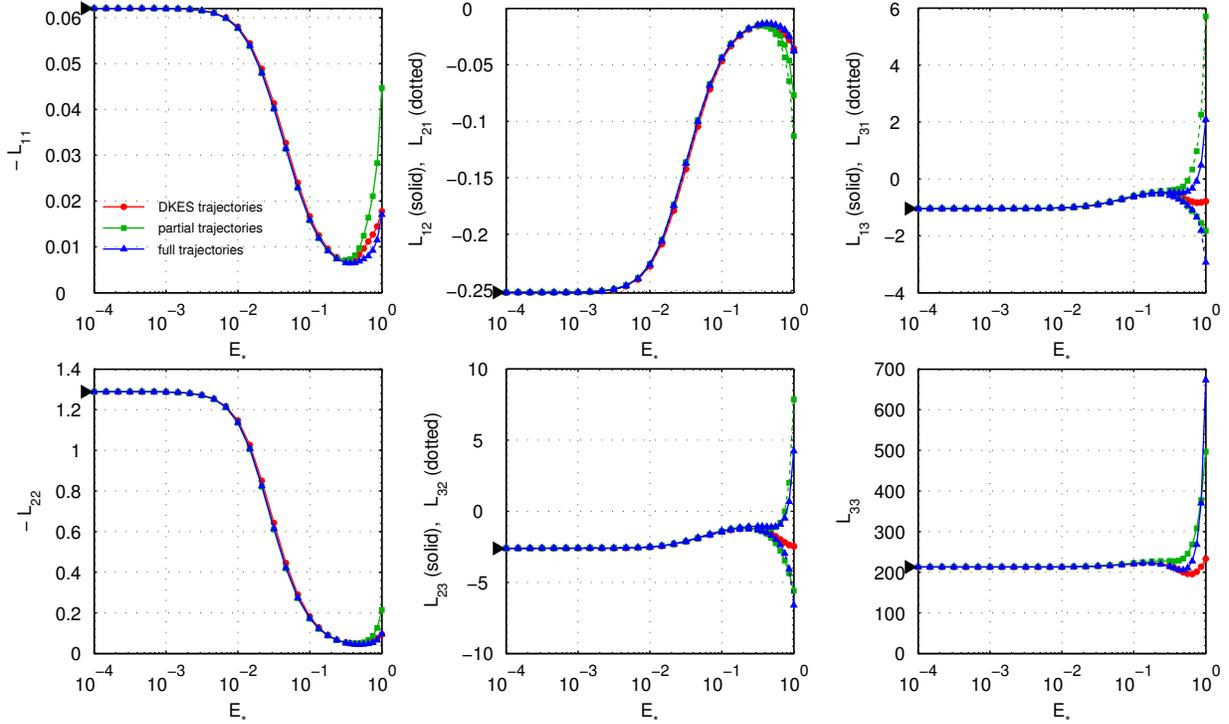}
\caption{(Color online) Comparison of trajectory models for LHD standard geometry at $\nu' = 0.01$,
using linearized Fokker-Planck collisions.
The ion transport matrix elements (defined in (\ref{eq:matrix})) are plotted as functions
of the normalized radial electric field (\ref{eq:EStar}).
Results for $E_r=0$ are indicated by the $\blacktriangleright$ symbol to the left of each plot.
\label{fig:ErComparison_LHD}}
\end{figure}

\begin{figure}[h!]
\includegraphics{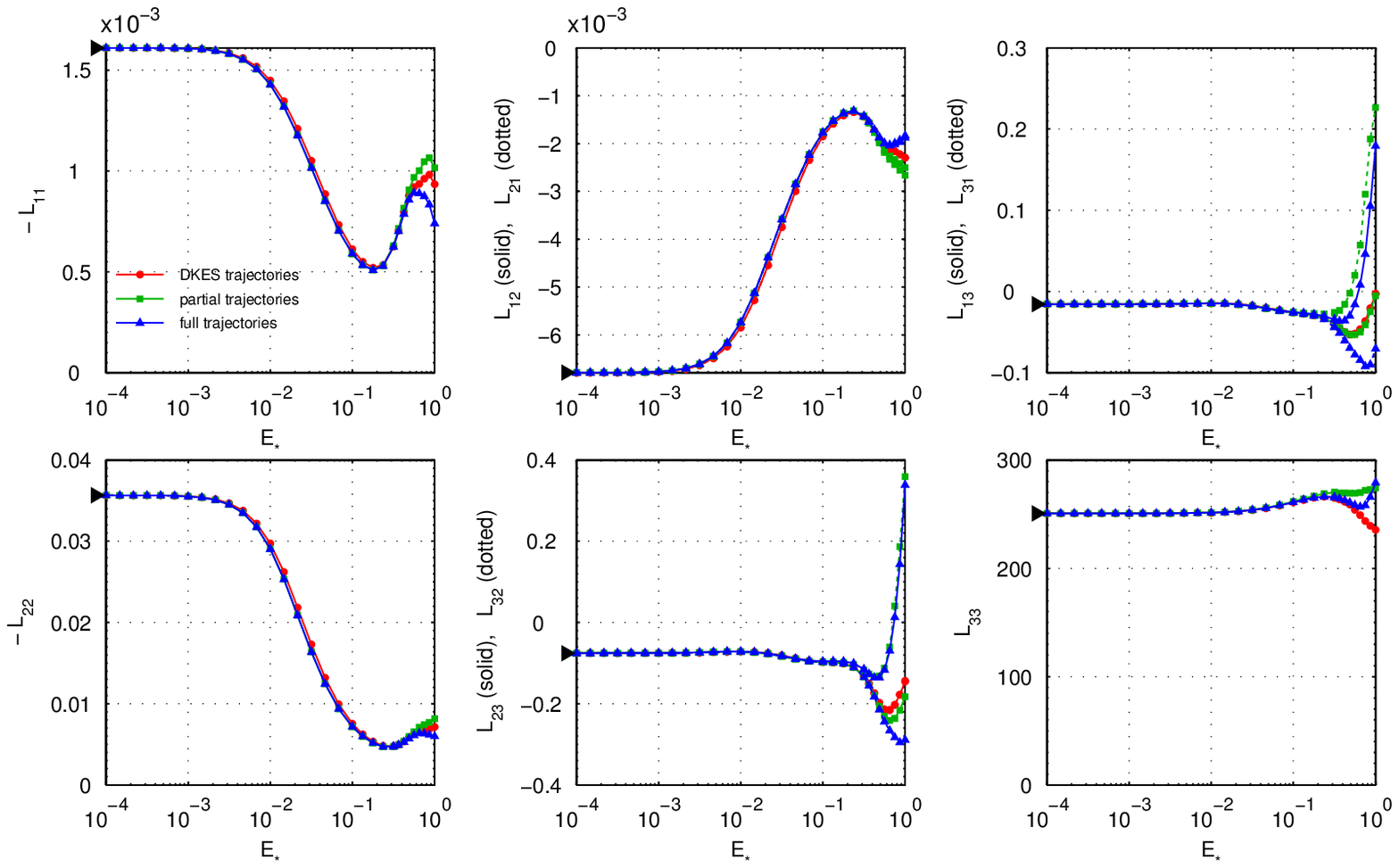}
\caption{(Color online) Comparison of trajectory models for W7-X standard geometry at $\nu' = 0.01$,
using linearized Fokker-Planck collisions.
The ion transport matrix elements (defined in (\ref{eq:matrix})) are plotted as functions
of the normalized radial electric field (\ref{eq:EStar}).
Results for $E_r=0$ are indicated by the $\blacktriangleright$ symbol to the left of each plot.
\label{fig:ErComparison_W7X}}
\end{figure}

Figures (\ref{fig:ErComparison_LHD}) and (\ref{fig:ErComparison_W7X}) show a SFINCS computation of
the ion transport matrix elements for the 3 trajectory models
in two different stellarator geometries.
The calculations in figure (\ref{fig:ErComparison_LHD}) are performed for the $r/a=0.5$ surface of
the LHD stellarator \cite{LHD} in its standard configuration. (Here the flux function $r$ is defined to be proportional to the square root of the toroidal flux enclosed by the flux surface in question.)
The calculations in figure (\ref{fig:ErComparison_W7X}) are performed for the $r/a=0.5$ surface of
the W7-X stellarator \cite{W7X1,W7X2} in its standard configuration.
In the LHD calculation, only the Boozer harmonics of $B(\theta,\zeta)/B_0$ with amplitude $>10^{-2}$
are retained, as listed in Table 1 of Ref. \cite{BeidlerBigBenchmarking},
whereas all harmonics with relative amplitude $>4\times 10^{-5}$ are retained for the W7-X calculations.
For both figures, the Fokker-Planck collision operator is used, and the collisionality
is set to $\nu'=0.01$.
As both figures illustrate, the electric field has negligible
effect on the transport matrix elements when $E_* < 0.01$.
For these small values of the electric field, the radial step size for diffusion is 
limited by collisions rather than by $\vect{E}\times\vect{B}$ precession.
As $E_r \to 0$, all the matrix elements converge smoothly to their $E_r=0$ limits.
For $E_*$ in the range $[0.01, 0.3]$,
the $\vect{E}\times\vect{B}$ precession suppresses radial transport,
as can be seen by the reduction in $|L_{11}|$ and $|L_{22}|$. In this regime of $E_*$, the three trajectory models give nearly
identical results for all the transport matrix elements.
However, once $E_*$ exceeds about 0.3, the results from the three trajectory models
begin to separate. 

In all probability, the reason why the three trajectory models agree
so well with each other below the resonance is that they all capture
the principal mechanism of transport in the $\sqrt{\nu}$-regime. The
$\vect{E}\times\vect{B}$ drift convects most locally trapped particles
poloidally around the torus, thus preventing them from drifting to the
wall, and the transport is instead dominated by shallowly trapped and
barely passing particles that are scattered back and forth across the
trapped-passing boundary on a time scale equal to the poloidal
convection time \cite{HoKulsrud}. This behavior is not likely to be
affected by the approximations 
made in the DKES and partial trajectory models.

In the typical ``ion root'' scenario, $E_*$ can be estimated by noting that the
ambipolar electric field arises to bring the ion particle transport
down to the electron level, and is therefore approximately determined so 
as to reduce the magnitude of the thermodynamic
force appearing as the first component of the vector on the
right-hand side of (\ref{eq:matrix}). The electric field is thus of
order $E_r \sim T/(eL_\perp)$, where $L_\perp$ denotes the length
scale corresponding to the pressure gradient. It is thus expected that $E_*$
is of order $E_* \sim \rho_\theta / L_\perp$, where $\rho_{\theta} =
\rho / (\iota \epsilon)$ and $\epsilon$ is the inverse aspect ratio,
and the ratio $\rho_\theta / L_\perp$ is typically $\ll 1$.
In W7-X, $E_*$ is predicted to be a few percent in normal plasma scenarios \cite{Turkin}.
The largest $E_r$ in normal W7-X scenarios is predicted to be a few tens of kV/m, in the
edge where density gradients are steep,
corresponding to $E_*$ up to a few tenths\cite{Turkin}.
However, in other previous experiments, scenarios with strong electron heating
can cause $T_e \gg T_i$,
giving rise to large positive ``electron root'' electric fields \cite{YokoyamaCERC}.
In these scenarios, $E_*$ may be $\sim 1$.

\begin{figure}[h!]
\includegraphics[height=7.5in]{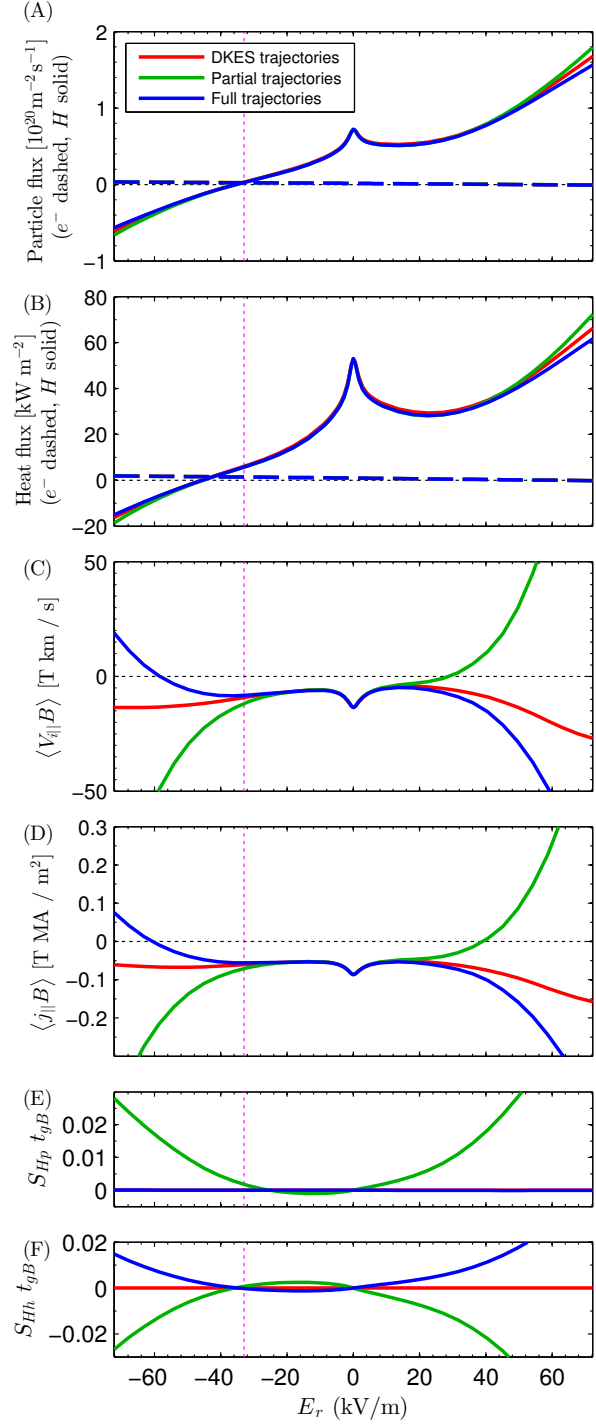}
\caption{(Color online)
Fluxes (A)-(B), flow (C), and bootstrap current (D)
computed for a scenario of steep pressure gradient near the edge of W7-X.
Magenta dotted line is the ambipolar $E_r$, effectively identical for the three
trajectory models.
The ion particle and heat sources in (E)-(F) are normalized
by a gyro-Bohm transport time $t_{gB}$.
\label{fig:2species}}
\end{figure}

Further analysis of whether the choice of trajectory model is significant in W7-X
is shown in figure  \ref{fig:2species}.
This calculation is based on the scenario considered in figure 5 of Ref. \cite{Turkin}. 
We focus on the radial location $r=0.45$ m ($r/a=0.88$)
in which the pressure gradient is strong.  This gradient should result in a large $E_r$, as predicted 
both by the argument in the preceding paragraph, and by
the modeling in Ref. \cite{Turkin} based on incompressible-$\vect{E}\times\vect{B}$ computations.
(Here, the flux label $r$ is defined by $\pi r^2 B_0 = 2 \pi \psi$.)
Matching the parameters in that work, we consider a pure hydrogen plasma with $n=6.6\times 10^{19}\; \mathrm{m}^{-3}$,
$\Te=\Ti=1$ keV, $dn/dr = -1.2\times 10^{21} \;\mathrm{m}^{-4}$, and $d\Te/dr=d\Ti/dr = -16$ keV/m.
These values correspond to $\nu'=0.03$ and $\Erres = 100 $ kV/m.
For this scenario, kinetic electrons are included in SFINCS along with the ions.
Inter-species linearized Fokker-Planck collisions are included with no expansion in mass ratio.

The radial fluxes of ions and electrons as functions of $E_r$ are shown in figure \ref{fig:2species}.A.
The electron fluxes (dashed curves) are very small $(\sim \sqrt{\me/\mi})$ compared to the ion fluxes
and are identical between the three trajectory models. No difference between the models is expected for
the electrons, since $E_*$ defined using the electron rather than ion thermal speed is always $\ll 1$.
The vertical magenta dotted line indicates the ambipolar value of $E_r \approx -33$  kV/m, which is effectively
identical for the three trajectory models, and comparable to the value predicted in \cite{Turkin}.
This electric field is roughly one third of the resonant value, and therefore the ion transport coefficients are
just beginning to separate for the three models. Heat fluxes are shown in figure \ref{fig:2species}.B, showing similar
behavior to the particle fluxes.

Figure \ref{fig:2species}.C shows the surface-averaged ion parallel flow. At the ambipolar value of $E_r$,
the three trajectory models yield similar values for the predicted flow.
At lower magnitudes of $E_r$, the flows predicted by the three models are nearly indistinguishable.
However, at larger electric fields, the three models begin to give quite different predictions.
This change in behavior around $E_* \sim 0.3$ is consistent with the patterns in figures \ref{fig:ErComparison_LHD}- \ref{fig:ErComparison_W7X}.
A similar pattern is visible in the bootstrap current density, shown in figure \ref{fig:2species}.D.
At the ambipolar value of $E_r$, the partial trajectory model predicts
$27\%$ more bootstrap current than the full trajectory model,
and the DKES model predicts $8\%$ more bootstrap current than the full trajectory model.
Interestingly, if the electric field exceeds 60 kV/m in the inward (ion root) direction, the bootstrap current
in the full trajectory model changes sign, whereas there is no sign change in the DKES model.

Figures (\ref{fig:2species}).E-F illustrate the ion particle and heat sources computed as part of the calculation.
As expected, the particle and heat sources are zero for the DKES model,
and for the full trajectory model, 
the particle source is always zero and the heat source vanishes at the ambipolar $E_r$.
Electron sources are negligible.
The plots show $S_{H\mathrm{p}}$ and $S_{H\mathrm{h}}$ from (\ref{eq:sources}) normalized to a gyro-Bohm transport time scale
$t_{gB} = L^2/D_{gB}$ with $D_{gB} = (\rho_i / L) cT/(eB_0)$, thereby
roughly normalizing the numerical sources to the scale of real physical sources arising from 
the divergence of the turbulent and neoclassical fluxes.
For this comparison we choose $L = -n / (dn/dr)$ to be the density scale length.
For the range of electric fields considered, the numerical sources are small on this transport time scale,
giving confidence in the model.
For the parameters considered, $\nuii t_{gB} = 0.4$, 
so dividing the values in figures (\ref{fig:2species}).E-F by this factor,
the source terms evidently remain much smaller
than the collision term in the kinetic equation for this calculation.

\section{Comparison of collision operators}
\label{sec:collisionComparison}

\begin{figure}[h!]
\includegraphics{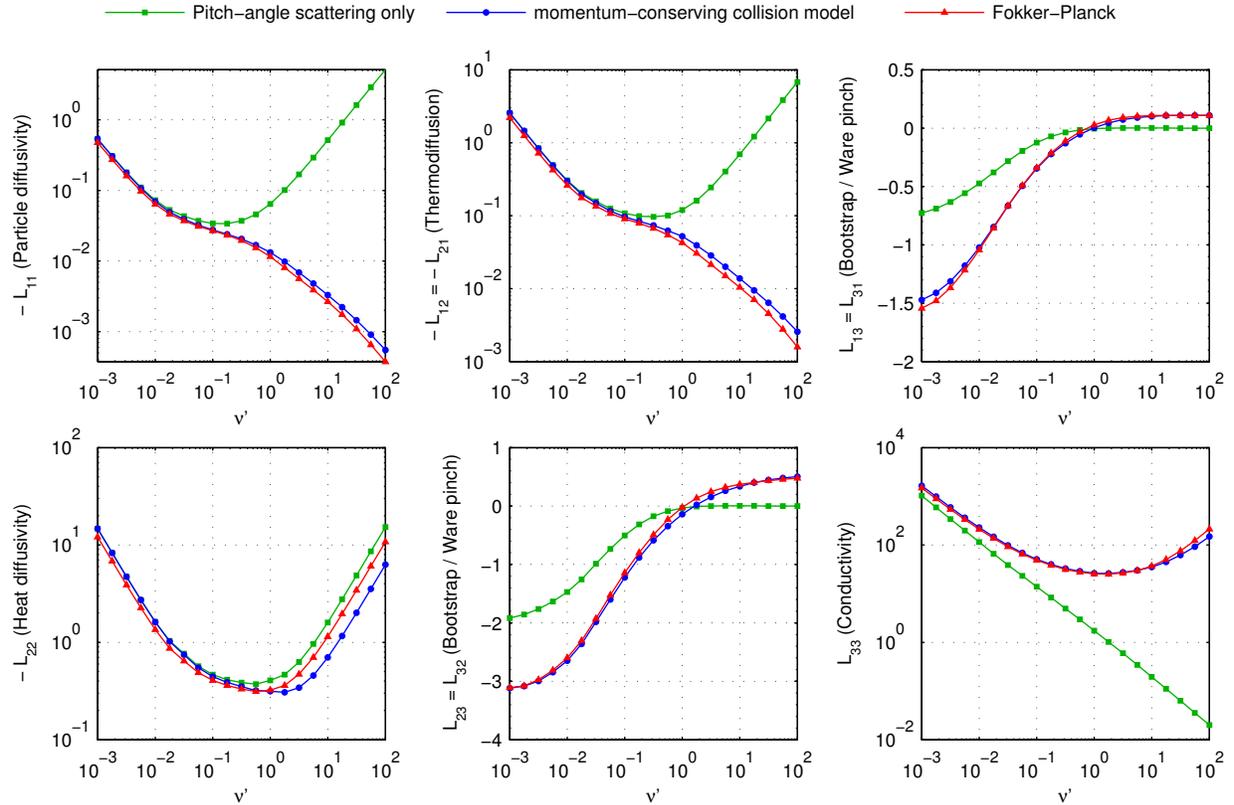}
\caption{(Color online) 
The ion transport matrix elements (defined in (\ref{eq:matrix})) are plotted as functions
of the collisionality (\ref{eq:nuPrime}) for LHD geometry at $E_r=0$.
SFINCS computations for three different collision operators are compared.
\label{fig:collisionComparison_LHD}}
\end{figure}

\begin{figure}[h!]
\includegraphics{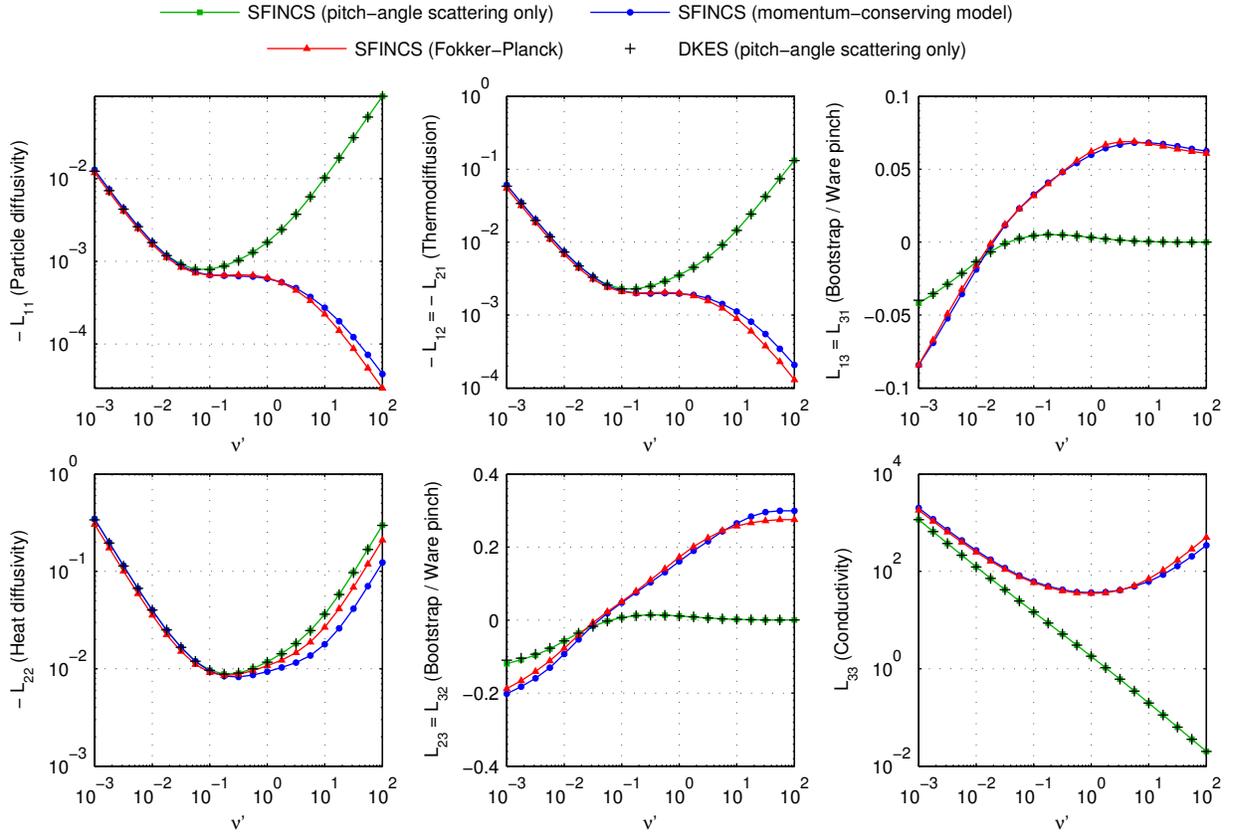}
\caption{(Color online)
The ion transport matrix elements (defined in (\ref{eq:matrix})) are plotted as functions
of the collisionality (\ref{eq:nuPrime}) for W7-X geometry at $E_r=0$.
SFINCS computations for three different collision operators are compared.
Also shown (black crosses) are the transport matrix elements
computed using the DKES code \cite{DKES1,DKES2}, which uses a pitch-angle scattering collision operator,
demonstrating excellent agreement with SFINCS when the latter is run with the same collision model. 
\label{fig:collisionComparison_W7X}}
\end{figure}

Figures (\ref{fig:collisionComparison_LHD})-(\ref{fig:collisionComparison_W7X}) show the transport
matrix elements for the LHD and W7-X geometries described earlier, this time comparing
the different collision operators as a function of collisionality.
The comparison is done for $d\Phi_0/d\psi=0$, so the three trajectory models
become identical, and the sources $S_a$ vanish.
It can be seen in the figures that at high collisionality,
momentum conservation is important for all the transport matrix elements (with the possible exception of $L_{22}$.)
At low collisionality, momentum conservation 
is unimportant for $L_{11}$, $L_{12}$, $L_{21}$, and $L_{22}$.
These matrix elements represent $1/\nu$-regime radial transport
(when $\nu' \ll 1$), which is associated with pitch-angle scattering of helically trapped particles.
Thus, the pitch-angle scattering approximation for collisions accurately
captures the dominant physics in these cases. 
When $d\Phi_0/d\psi \ne 0$, the same is true for the
$\sqrt{\nu}$-regime, where the main effect of the collisions is to
scatter particles across a thin collisional boundary layer in velocity
space around the trapped-passing boundary.

The other matrix elements ($L_{13}$, $L_{23}$, $L_{31}$, $L_{32}$, and $L_{33}$) 
are more sensitive to momentum conservation at low collisionality.
For all the matrix elements at all collisionalities, the momentum-conserving model operator
reproduces all the trends of the more accurate linearized Fokker-Planck operator,
though with some $O(1)$ differences.

Note in Figs.~(\ref{fig:collisionComparison_LHD})-(\ref{fig:collisionComparison_W7X}) that the scaling of the $L_{11}$ and $L_{12}$ coefficients at high collisionality depends crucially on whether momentum is conserved in the collision operator. In the momentum-conserving calculations, these transport coefficients are inversely proportional to $\nu$, whereas they are proportional to $\nu$ if the collisions are approximated by pure pitch-angle scattering. To understand why, it is useful to recall that the Pfirsch-Schl{\"u}ter
particle flux consists of two terms: one related to the parallel
friction force and one related to parallel viscosity \cite{AndreiPer2009, Braun}. This is most easily seen by taking the scalar product of the lowest-order plasma current, which satisfies $\vect{J} \times \vect{B} = c p'(\psi) \nabla \psi$, with the momentum equation,
\begin{equation}
m_a n_a \vect{V}_a \cdot \nabla \vect{V}_a = n_a e_a \left( - \nabla \Phi + c^{-1}\vect{V}_a \times \vect{B} \right)
- \nabla p_a - \nabla \cdot \tens{\pi}_a + \vect{F}_a, 
\label{eq:vectorMomentum}
\end{equation}
neglecting the left-hand side. Since $\nabla \cdot \vect{J} = 0$ and $n_a$ is a flux function in lowest order, we obtain
\begin{equation}
  \left< n_a \vect{V}_a \cdot \nabla \psi \right> 
  = \frac{1}{e_a p'(\psi)} \left< \vect{J} \cdot \left( \vect{F}_a - \nabla \cdot \tens{\pi}_a \right) \right>,
\end{equation}
where the term corresponding to the perpendicular component of the friction force $\vect{F}_a$ represents the classical particle flux and the other terms the neoclassical flux,
\begin{equation}
  \left< n_a \vect{V}_a \cdot \nabla \psi \right>_{nc} = \frac{1}{e_ap'} \left< J_\| F_{a\|} + \tens{\pi}_a:\nabla \vect{J} \right>
  \label{PS flux}
\end{equation}
where the viscosity tensor is $\tens{\pi}_a =(p_{a\|} - p_{a\perp})(\vect{b b} - \tens{I}/3)$.
The first term in (\ref{PS flux}) is proportional to $\nu$ and
therefore dominates at high collisionality, but vanishes when there is only a 
single ion species because of momentum conservation in like-particle
collisions. All that remains is therefore the particle flux caused
by parallel viscosity, which is inversely proportional to $\nu$ at high collisionality \cite{PerBook}. In
the pure pitch-angle-scattering model however, parallel momentum
conservation is violated, leading to spurious friction-driven
transport proportional to $\nu$. This is why the green curves have a
slope of +1 for large $\nu$ in the logarithmic plots of $L_{11}$ and
$L_{12}$ in Figures
(\ref{fig:collisionComparison_LHD})-(\ref{fig:collisionComparison_W7X}),
while the blue and red curves have the slope -1.

A similar difference between the momentum-conserving and
pitch-angle-scattering operators is evident in the parallel
conductivity coefficient $L_{33}$. The flow that arises in response to
a parallel electric field is determined by the parallel momentum equation $\vect{b}\cdot$(\ref{eq:vectorMomentum}),
where the parallel friction force $F_{\|a}$ again vanishes when only a
single ion species is considered. Hence 
\begin{equation} 
n_a Z_a e \left< B  E_{||} \right> = \left< \vect{B}\cdot(\nabla \cdot \tens{\pi}_a) \right> 
= \left< (p_\perp - p_\| ) \nabla_\| B \right>.
\label{ParallelMomentum}
\end{equation}
In
the absence of radial gradients, the pressure anisotropy in the \PS
regime is proportional to the parallel flow velocity and inversely
proportional to the collision frequency \cite{AndreiPer2009}. The
flow $\left< V_\| B \right>$ is therefore proportional to $\nu$ in
the \PS regime unless momentum conservation is violated. In the
latter case, the spurious friction force causes $\left< V_\| B
\right>$ to be inversely proportional to $\nu$, as can be seen in
Figs.~(\ref{fig:collisionComparison_LHD})-(\ref{fig:collisionComparison_W7X}).

When $\nu' < 1$, the resolution required in the $\theta$, $\zeta$, and $\xi$ coordinates
increases as $\nu'$ decreases,
due to the boundary layers that develop in phase space.  The highest resolution used
for results presented here, corresponding to the W7-X calculations at  $\nu'=10^{-3}$,
was $N_\theta=29$, $N_\zeta = 83$, $N_\xi=180$, and $N_x=5$,
giving a $2,166,302 \times 2,166,302$ linear system.
Here, $N_j$ is the number of grid points or modes in coordinate $j$.
Each calculation at this resolution with the Fokker-Planck collision operator
required $\sim 30-50$ minutes to run on 4 nodes of the Edison computer at NERSC.
At higher collisionality, or if fewer harmonics are retained in $B(\theta,\zeta)$, 
lower resolution is sufficient, so memory and time requirements are reduced;
for example, in the same W7-X geometry at $\nu'=10^{-2}$, sufficient resolution parameters for convergence
were $N_\theta=11$, $N_\zeta=64$, $N_\xi=100$, and $N_x=5$,
and computations required 3 minutes on 1 node of Edison.
Computations with $\nu' > 10^{-2}$ can typically be run on a laptop.

When the pure pitch-angle scattering collision operator 
and DKES trajectories are chosen, the kinetic equation
solved in SFINCS becomes identical to the one solved in the DKES code \cite{DKES1, DKES2}.
In this case, it was verified that the two codes agreed for all elements of the transport
matrix, as demonstrated in figure \ref{fig:collisionComparison_W7X}.
For this figure, the monoenergetic transport coefficients
computed by DKES have been integrated over velocity with the appropriate
weights and normalized in the same way as (\ref{eq:matrix}).

In the short-mean-free-path limit $\nu' \gg 1$, the ion transport and flow can be computed
analytically in terms of the parallel current \cite{AndreiPer2009}.
The transport matrix elements associated with the Fokker-Planck collision operator
may therefore be extracted from Ref. \cite{AndreiPer2009} and are summarized in Appendix \ref{apdx:SH}.
Plotted in figure \ref{fig:SimakovHelanderComparison} (dashed and dot-dashed lines),
these analytic high-collisionality limits agree quite well with the Fokker-Planck SFINCS computations
in the appropriate $\nu' \gg 1$ limit.

\begin{figure}[h!]
\includegraphics{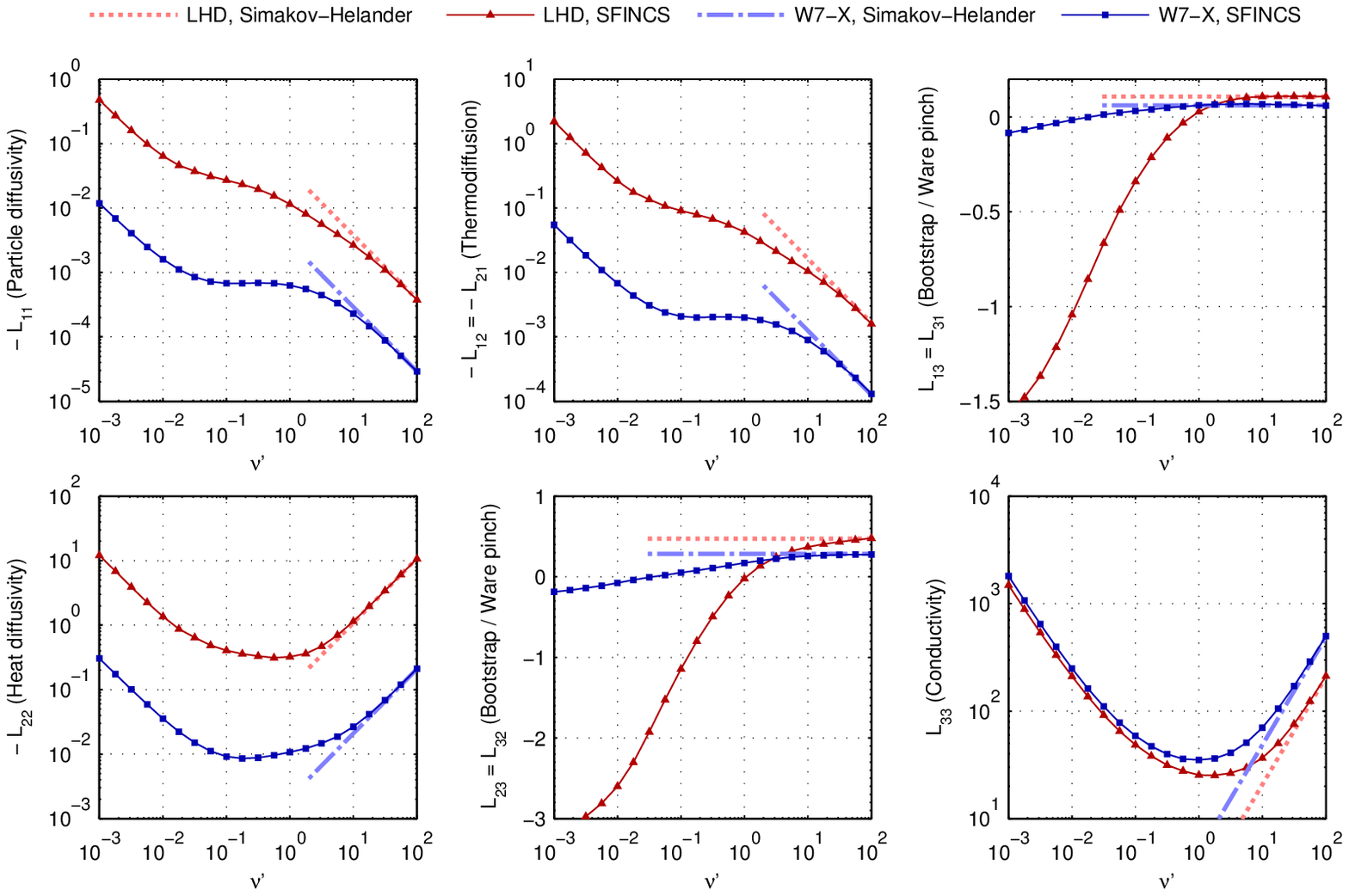}
\caption{(Color online)
The ion transport matrix elements (defined in (\ref{eq:matrix})) are plotted as functions
of the collisionality (\ref{eq:nuPrime}) for LHD and W7-X geometry at $E_r=0$.
SFINCS results shown were computed using the linearized Fokker-Planck collision operator
(so the solid curves here are
identical to the red curves in figures \ref{fig:collisionComparison_LHD}-\ref{fig:collisionComparison_W7X}.)
Dashed and dot-dashed lines indicate the analytic high-collisionality limits
for Fokker-Planck collisions, discussed in Ref. \cite{AndreiPer2009} and in appendix \ref{apdx:SH},
which agree quite well with the SFINCS computations at high collisionality.
\label{fig:SimakovHelanderComparison}}
\end{figure}

\section{Discussion and conclusions}
\label{sec:conclusions}

In this work, we have examined the impact of several approximations made in 
stellarator kinetic codes, approximations related to the electric field and the collision operator.
We have compared three versions of the drift-kinetic equation
for a stellarator, consisting of (\ref{eq:kinetic4}) with
the coefficients (\ref{eq:fullTrajectories}), (\ref{eq:DKESTrajectories}),
or (\ref{eq:partialTrajectories}).
These three sets of expressions for $\vect{\dot{r}}_a$,
$\dot{x}_a$, and $\dot{\xi}_a$ may be interpreted as effective
particle trajectories (although we solve each form of the kinetic equation
using continuum numerical methods).
Equations  (\ref{eq:fullTrajectories}) and  (\ref{eq:partialTrajectories})
appear to be more accurate than  (\ref{eq:DKESTrajectories}),
and as we have shown in section \ref{sec:momentum},
the full trajectory model (\ref{eq:fullTrajectories}) is the only
one of the three models which gives the correct parallel viscous force 
and which preserves intrinsic ambipolarity in quasisymmetry.
However, as we have shown in section \ref{sec:sources},
the kinetic equation (\ref{eq:kinetic4})
with  (\ref{eq:fullTrajectories}) or  (\ref{eq:partialTrajectories})
is not well behaved when $E_r \ne 0$,
with one or two unphysical constraints forced upon the distribution function.
This analytic property of the kinetic equation must be dealt with
before attempting to solve the equation numerically.

To eliminate this problem of unphysical constraints,
we propose formulating the kinetic problem as in  (\ref{eq:blocks}) with (\ref{eq:sources}).
A particle and heat source are introduced,
along with the additional constraints that all the flux-surface-averaged
density and pressure reside in the leading-order Maxwellian.
For the model (\ref{eq:DKESTrajectories}), the sources always vanish.
For the model  (\ref{eq:fullTrajectories}), the particle source
vanishes for any $E_r$, and the energy source vanishes when $E_r$
takes on its ambipolar value.
The equations (\ref{eq:blocks}) have been implemented in a new time-independent continuum code SFINCS,
and the resulting ion transport matrices have been compared for the geometries of the LHD and W7-X stellarators.
When $E_r$ is below roughly one-third of the resonant value,
the three models give nearly indistinguishable results. 
This finding confirms that the incompressible-$\vect{E}\times\vect{B}$ trajectory
model used in some codes \cite{DKES1,DKES2} is quite accurate in this
small-$E_r$ regime, which is typically satisfied in experiments.
Physically, the effect of $E_r$ in this regime is to generate a $\sqrt{\nu}$ regime
of transport due to poloidal precession of helically trapped particles,
and this process is retained (at least approximately) in all three trajectory models.
Once $E_r$ approaches the resonance, however, the three trajectory models
yield substantially different results.
This $E_r \sim \Erres$ regime can be relevant to experiments with high ratios $\Te / \Ti$
\cite{YokoyamaCERC, Lore, Briesemeister} and strong gradients \cite{Baldzuhn}.
In figure \ref{fig:2species}, we find that in the large-$E_r$ region anticipated for the edge of W7-X,
the bootstrap current density in the full trajectory model is modestly reduced (by $8\%$)
compared to an incompressible-$\vect{E}\times\vect{B}$ calculation,
but should larger values of $E_*$ arise, we expect the deviation could grow more significant.

Since full coupling in the speed coordinate $x_{a}$ is retained in our numerical
implementation, it is possible to directly compare results from the full linearized
Fokker-Planck collision operator to results from simpler collision models.
At low collisionality, the ion transport matrix elements $L_{11}$,
$L_{12}$, $L_{21}$, and $L_{22}$ are nearly identical for the three collision models
considered. This result makes sense physically since these matrix elements at low collisionality
are associated with a piece of the distribution function that is localized to a narrow range of pitch angles,
so pitch angle diffusion is the dominant collisional process.
However, these same matrix elements at higher collisionality, or the other matrix elements
at any collisionality, are sensitive to momentum conservation.
The momentum-conserving model operator results in the correct scaling with collisionality
when compared to the full linearized Fokker-Planck operator.
However, there are still $O(1)$ differences in the transport coefficients
computed with these two collision operators.

\begin{acknowledgments}
This work was supported by the US Department of Energy through grants DE-FG02-91ER-54109 and DE-FG02-93ER-54197.
M. L. is grateful to the Plasma Science and Fusion Center at the Massachusetts Institute of Technology,
where he carried out much of the code development, and for travel support from the
Max-Planck-Institut f\"{u}r Plasmaphysik in Greifswald, Germany.
Some of the computer simulations presented here
used resources of the National Energy Research Scientific Computing Center (NERSC),
which is supported by the Office of Science of the U.S. Department of Energy under Contract No. DE-AC02-05CH11231.
M. L. was supported by the
Fusion Energy Postdoctoral Research Program
administered by the Oak Ridge Institute for Science and Education.
We are thankful to J. Geiger for providing the W7-X equilibrium data,
and to Craig Beidler and Peter Catto for other input on this work.
We also thank the anonymous referee for making several suggestions
that substantially improved the paper.

\end{acknowledgments}

\appendix
\section{Quasisymmetry Isomorphism}

A useful test of a stellarator neoclassical code such as the one
described here is the quasisymmetry isomorphism,
discussed analytically in Refs. \cite{Pytte, Boozer83, meQS}.
Equivalent to the definition at the end of section \ref{sec:momentum} \cite{PerAndreiPRL},
a quasisymmetric magnetic field is one satisfying $B(\theta,\zeta) = y(M\theta-N\zeta)$
for some periodic function $y$ and integers $M$ and $N$.
Magnetic fields with the same $y$ but different $M$ and $N$ are said to be isomorphic in that
the associated transport matrices must be related in the following manner.
Suppose the transport matrices are computed for several quasisymmetric magnetic fields with different values of $M$ and $N$,
varying the collision frequency in each calculation so
$\nuii / (\iota M - N)$ remains fixed,
and varying the radial electric field so
$(d\Phi_0/d\psi)(GM+IN)/(\iota M-N)$
remains fixed.
In such a scan of $M$ and $N$,
it can be shown analytically \cite{Pytte, Boozer83, meQS} that the 
transport matrix elements should vary as follows:
$L_{11}$, $L_{12}$, $L_{21}$, and $L_{22}$ vary
$\propto (NI+MG)^2/(\iota M-N)$;
$L_{13}$, $L_{23}$, $L_{31}$, and $L_{32}$ vary
$\propto (NI+MG)/(\iota M-N)$;
and $L_{33}$ varies
$\propto 1/(\iota M-N)$.
This isomorphism holds for all the trajectory models considered in this paper.

As $\nuii / (\iota M - N)$ is to be held fixed in this test, while $\iota M - N$ can change sign as $M$ and $N$ are varied, 
the collision frequency to use can be negative.
While $\nuii <0$ does not make sense physically, it poses no mathematical or numerical problem.
In any stellarator (even a non-quasisymmetric and/or non-stellarator-symmetric one),
if the signs of the collision frequency, $\xi$, and $d\Phi_0/d\psi$ are simultaneously
reversed in the kinetic equation, the part of $f_{i1}$ driven by $\left< E_{||}B\right>$ remains unchanged,
while the part driven by radial gradients changes sign. Thus, $L_{11}$, $L_{12}$, $L_{21}$, $L_{22}$, and $L_{33}$ change sign, while
$L_{13}$, $L_{23}$, $L_{31}$, and $L_{32}$ remain unchanged.
Therefore, another way to express the quasisymmetry isomorphism 
(even for non-stellarator-symmetric $y$)
that preserves $\nuii > 0$ is the following:
if $M$ and $N$ are varied holding 
$\nuii / \left|\iota M - N\right|$ and $(d\Phi_0/d\psi)(GM+IN)/|\iota M-N|$ fixed,
$L_{11}$, $L_{12}$, $L_{21}$, and $L_{22}$ should vary
$\propto (NI+MG)^2/|\iota M-N|$;
$L_{13}$, $L_{23}$, $L_{31}$, and $L_{32}$ should vary
$\propto (NI+MG)/(\iota M-N)$;
and $L_{33}$ should vary
$\propto 1/|\iota M-N|$.

It was verified that the SFINCS code obeyed both versions of these isomorphism transformations
for various $y$, collisionality regimes, radial electric fields, trajectory models, and collision operators.


\section{Ion transport matrix at high collisionality}
\label{apdx:SH}

From the analytic calculations presented in Ref.~\cite{AndreiPer2009} we can derive expressions
for $L_{jk}$ of Eq.~(\ref{eq:matrix}) in the Pfirsch-Schl{\"u}ter regime. 
\changed{Note that a pure plasma with singly charged ions is assumed in Ref. \cite{AndreiPer2009}, so we specialize to this case of $Z=1$ in this appendix.}
The transport matrix elements depend on the function $u$ given by the solution to Eq.~(8) in Ref.~\cite{AndreiPer2009};
$u$ is proportional to the parallel current divided by $B$. 
All coefficients but $L_{33}$ are straightforwardly obtained from Eqs.~(14), (18) and (26) in Ref.~\cite{AndreiPer2009} for the radial ion heat flux, the parallel ion flow, and the radial current respectively, by 
suitable choices of 
the thermodynamic forces in the right-hand-side of Eq.~(\ref{eq:matrix}) and using the symmetry of the transport matrix.
To find the parallel conductivity coefficient $L_{33}$ we substitute the pressure anisotropy, given by Eq.~(20) in Ref.~\cite{AndreiPer2009}, into Eq.~(\ref{ParallelMomentum}) (of this paper) 
in the absence of radial gradients (i.e. when $E_{||}$ is the only thermodynamic force present). Then $L_{33}$ can be found from the flow $\left< V_\| B \right>$.

Expressions for the matrix coefficients in the Pfirsch-Schl{\"u}ter regime are summarized in 
Eqs.~(\ref{eq:LPSregime})-(\ref{eq:PSregimeFunctions}). 
Three numerical coefficients in the function ${K}_2^{\mathrm{Simakov}} \left(\psi\right)$
arise from generalized Spitzer problems, 
which were solved in Ref.~\cite{AndreiPer2009} by keeping a small number of Laguerre polynomials in kinetic energy.
When these generalized Spitzer problems are solved keeping more energy polynomials,
we obtain the more accurate coefficients given in ${K}_{2}$ below.

\begin{align}
\begin{split}
L_{11} & = 
 0.96 \cdot 2^{1/2} \cdot  \frac{3}{4}  \frac{\left(G + \iota I\right)^2}{ \iota^2 G^2} 
 {G}_1 \left(\psi\right) \frac{1}{\nu'},\\
L_{12} & = L_{21} =  0.96 \cdot 2^{1/2} \frac{\left({G} + \iota {I}\right)^2}{ \iota^2 {G}^2} \left[ 3.245 \, {G}_1 \left(\psi\right) + 0.085 \, {G}_2 \left(\psi\right)  \right] \frac{1}{\nu'}, \\
L_{13} & = L_{31}  = \frac{\left\langle {u} {B}^2  \right\rangle}{{G} \iota} - \frac{\left\langle {B}^2\right\rangle}{{G} \iota} {K}_1\left(\psi\right), \\
L_{22} & = 2^{1/2} \cdot  \frac{8}{5} \frac{1}{\iota^2 {G}^2} B_0^2 {H} \left(\psi\right) \, \nu', \\
L_{23} & = L_{32} = \frac{1}{\iota {G}} \left[\frac{5}{2} \left\langle {u} {B}^2  \right\rangle - \frac{5}{2} {K}_1\left(\psi\right) \left\langle {B}^2\right\rangle + {K}_2\left(\psi\right) \left\langle {B}^2\right\rangle\right], \\
L_{33} & = \frac{1}{3 \cdot 0.96 \cdot 2^{1/2}} \frac{1}{\left({G} + \iota {I}\right)^2} \frac{\left\langle {B}^2\right\rangle^2}{\left\langle \left( {\nabla}_\| {B}\right)^2\right\rangle} \, \nu',	
\label{eq:LPSregime}
\end{split}
\end{align}
\begin{align}
\begin{split}
& {G}_1 \left(\psi\right) 
= \frac{\left\langle \left( {\nabla}_\| \ln {B}\right)  {\nabla}_\| \left({u} {B}^2\right)\right\rangle^2}{\left\langle \left( {\nabla}_\| {B}\right)^2\right\rangle} -
 \left\langle \left[\frac{ {\nabla}_\| \left({u} {B}^2\right)}{{B}}\right]^2\right\rangle,  \\
& {G}_2 \left(\psi\right) 
 = \left\langle {u} \left({\nabla}_\| \ln {B}\right) {\nabla}_\| \left({u} {B}^2\right) \right\rangle
 - \frac{\left\langle \left( {\nabla}_\| \ln {B}\right)  {\nabla}_\| \left({u} {B}^2\right)\right\rangle
 \left\langle {u} \left( {\nabla}_\| {B}\right)^2\right\rangle}{\left\langle \left( {\nabla}_\| {B}\right)^2\right\rangle},  \\
& {K}_1 \left(\psi\right) 
 = \frac{\left\langle \left( {\nabla}_\| \ln {B}\right)  {\nabla}_\| \left({u} {B}^2\right)\right\rangle}{2 \left\langle \left( {\nabla}_\| {B}\right)^2\right\rangle}, \\
& {K}_2 \left(\psi\right) 
 = 1.97213 \frac{\left\langle {u} {B}^2  \right\rangle}{\left\langle {B}^2  \right\rangle} - 1.03287 \cdot 2 {K}_1 \left(\psi\right)
 + 0.09361 \frac{
 \left\langle {u} \left( {\nabla}_\| {B}\right)^2\right\rangle}{\left\langle \left( {\nabla}_\| {B}\right)^2\right\rangle}, \\
&  {K}_{2}^{\mathrm{Simakov}} \left(\psi\right) 
  = 1.77 \frac{\left\langle {u} {B}^2  \right\rangle}{\left\langle {B}^2  \right\rangle} - 0.91 \cdot 2 {K}_1 \left(\psi\right)
 + 0.05 \frac{
 \left\langle {u} \left( {\nabla}_\| {B}\right)^2\right\rangle}{\left\langle \left( {\nabla}_\| {B}\right)^2\right\rangle}, \\
& {H} \left(\psi\right) 
 = \frac{\left\langle {u} {B}^2  \right\rangle^2}{\left\langle {B}^2  \right\rangle} - \left\langle {u}^2 {B}^2  \right\rangle.
\label{eq:PSregimeFunctions}
\end{split}
\end{align}

\bibliography{sfincsPaper}

\end{document}